\pdfoutput=1

\documentclass[12pt,a4paper]{article}

\usepackage[skip=4pt]{caption}

\usepackage{ifthen} 
\newboolean{pdflatex}
\setboolean{pdflatex}{true} 

\newboolean{articletitles}
\setboolean{articletitles}{true} 

\newboolean{uprightparticles}
\setboolean{uprightparticles}{false} 

\newboolean{inbibliography}
\setboolean{inbibliography}{false} 

\usepackage[top=1in, bottom=1.25in, left=1in, right=1in]{geometry}

\usepackage{microtype}
\usepackage{lineno}  
\usepackage{xspace} 
\usepackage{caption} 

\usepackage{graphicx}  
\usepackage{color}
\usepackage{colortbl}
\graphicspath{{./figs/}} 

\usepackage{amsmath} 
\usepackage{amssymb}
\usepackage{amsfonts}
\usepackage{upgreek} 

\newcommand*\patchAmsMathEnvironmentForLineno[1]{%
\expandafter\let\csname old#1\expandafter\endcsname\csname #1\endcsname
\expandafter\let\csname oldend#1\expandafter\endcsname\csname
end#1\endcsname
 \renewenvironment{#1}%
   {\linenomath\csname old#1\endcsname}%
   {\csname oldend#1\endcsname\endlinenomath}%
}
\newcommand*\patchBothAmsMathEnvironmentsForLineno[1]{%
  \patchAmsMathEnvironmentForLineno{#1}%
  \patchAmsMathEnvironmentForLineno{#1*}%
}
\AtBeginDocument{%
\patchBothAmsMathEnvironmentsForLineno{equation}%
\patchBothAmsMathEnvironmentsForLineno{align}%
\patchBothAmsMathEnvironmentsForLineno{flalign}%
\patchBothAmsMathEnvironmentsForLineno{alignat}%
\patchBothAmsMathEnvironmentsForLineno{gather}%
\patchBothAmsMathEnvironmentsForLineno{multline}%
\patchBothAmsMathEnvironmentsForLineno{eqnarray}%
}

\usepackage{hyperref}    
\usepackage[all]{hypcap} 

\usepackage{xspace} 
\usepackage{upgreek}

\def\lhcb {\mbox{LHCb}\xspace}
\def\atlas  {\mbox{ATLAS}\xspace}
\def\cms    {\mbox{CMS}\xspace}

\def\babar  {\mbox{BaBar}\xspace}
\def\belle  {\mbox{Belle}\xspace}
\def\cleo   {\mbox{CLEO}\xspace}
\def\cdf    {\mbox{CDF}\xspace}

\def\delphi {\mbox{DELPHI}\xspace}

\def\lthree {\mbox{L3}\xspace}

\def\lhc    {\mbox{LHC}\xspace}
\def\lep    {\mbox{LEP}\xspace}

\def\MagUp {\mbox{\em Mag\kern -0.05em Up}\xspace}

\ifthenelse{\boolean{uprightparticles}}%
{

 \def\Peta        {\ensuremath{\upeta}\xspace}

 \def\Pmu         {\ensuremath{\upmu}\xspace}

 \def\Pchi        {\ensuremath{\upchi}\xspace}                 
 \def\Ppsi        {\ensuremath{\uppsi}\xspace}

 \def\PDelta      {\ensuremath{\Delta}\xspace}                 
 \def\PXi      {\ensuremath{\Xi}\xspace}                 
 \def\PLambda      {\ensuremath{\Lambda}\xspace}                 
 \def\PSigma      {\ensuremath{\Sigma}\xspace}                 
 \def\POmega      {\ensuremath{\Omega}\xspace}                 
 \def\PUpsilon      {\ensuremath{\Upsilon}\xspace}                 
 
 \def\PB      {\ensuremath{\mathrm{B}}\xspace}                 
                  
 \def\PD      {\ensuremath{\mathrm{D}}\xspace}

 \def\PJ      {\ensuremath{\mathrm{J}}\xspace}                 
 \def\PK      {\ensuremath{\mathrm{K}}\xspace}

 \def\Pb      {\ensuremath{\mathrm{b}}\xspace}                 
 \def\Pc      {\ensuremath{\mathrm{c}}\xspace}                 
                  
 \def\Pe      {\ensuremath{\mathrm{e}}\xspace}

 \def\Pi      {\ensuremath{\mathrm{i}}\xspace}

 \def\Pp      {\ensuremath{\mathrm{p}}\xspace}

 \def\Ps      {\ensuremath{\mathrm{s}}\xspace}

}
{

 \def\Peta        {\ensuremath{\eta}\xspace}

 \def\Pmu         {\ensuremath{\mu}\xspace}

 \def\Pchi        {\ensuremath{\chi}\xspace}                 
 \def\Ppsi        {\ensuremath{\psi}\xspace}                 
                  
 \mathchardef\PDelta="7101
 \mathchardef\PXi="7104
 \mathchardef\PLambda="7103
 \mathchardef\PSigma="7106
 \mathchardef\POmega="710A
 \mathchardef\PUpsilon="7107
                  
 \def\PB      {\ensuremath{B}\xspace}                 
                  
 \def\PD      {\ensuremath{D}\xspace}

 \def\PJ      {\ensuremath{J}\xspace}                 
 \def\PK      {\ensuremath{K}\xspace}

 \def\Pb      {\ensuremath{b}\xspace}                 
 \def\Pc      {\ensuremath{c}\xspace}                 
                  
 \def\Pe      {\ensuremath{e}\xspace}

 \def\Pi      {\ensuremath{i}\xspace}

 \def\Pp      {\ensuremath{p}\xspace}

 \def\Ps      {\ensuremath{s}\xspace}

}

\makeatletter
\ifcase \@ptsize \relax
  \newcommand{\miniscule}{\@setfontsize\miniscule{4}{5}}
\or
  \newcommand{\miniscule}{\@setfontsize\miniscule{5}{6}}
\or
  \newcommand{\miniscule}{\@setfontsize\miniscule{5}{6}}
\fi
\makeatother

\DeclareRobustCommand{\optbar}[1]{\shortstack{{\miniscule (\rule[.5ex]{1.25em}{.18mm})}
  \\ [-.7ex] $#1$}}

\def\epem       {{\ensuremath{\Pe^+\Pe^-}}\xspace}

\def\mup        {{\ensuremath{\Pmu^+}}\xspace}
\def\mun        {{\ensuremath{\Pmu^-}}\xspace} 

\def\squark    {{\ensuremath{\Ps}}\xspace}
\def\squarkbar {{\ensuremath{\overline \squark}}\xspace}
\def\ssbar     {{\ensuremath{\squark\squarkbar}}\xspace}
\def\cquark    {{\ensuremath{\Pc}}\xspace}
\def\cquarkbar {{\ensuremath{\overline \cquark}}\xspace}
\def\ccbar     {{\ensuremath{\cquark\cquarkbar}}\xspace}
\def\bquark    {{\ensuremath{\Pb}}\xspace}

\def\kaon    {{\ensuremath{\PK}}\xspace}
  \def\Kbar    {{\kern 0.2em\overline{\kern -0.2em \PK}{}}\xspace}

\def\KorKbar    {\kern 0.18em\optbar{\kern -0.18em K}{}\xspace}

\def\Kp      {{\ensuremath{\kaon^+}}\xspace}
\def\Km      {{\ensuremath{\kaon^-}}\xspace}

\def\KS      {{\ensuremath{\kaon^0_{\mathrm{ \scriptscriptstyle S}}}}\xspace}

  \def\Dbar    {{\kern 0.2em\overline{\kern -0.2em \PD}{}}\xspace}

\def\DorDbar    {\kern 0.18em\optbar{\kern -0.18em D}{}\xspace}

\def\B       {{\ensuremath{\PB}}\xspace}
\def\Bbar    {{\ensuremath{\kern 0.18em\overline{\kern -0.18em \PB}{}}}\xspace}

\def\BorBbar    {\kern 0.18em\optbar{\kern -0.18em B}{}\xspace}
\def\Bz      {{\ensuremath{\B^0}}\xspace}

\def\Bu      {{\ensuremath{\B^+}}\xspace}

\def\Bp      {{\ensuremath{\Bu}}\xspace}

\def\Bs      {{\ensuremath{\B^0_\squark}}\xspace}

\def\Bcp     {{\ensuremath{\B_\cquark^+}}\xspace}

\def\jpsi     {{\ensuremath{{\PJ\mskip -3mu/\mskip -2mu\Ppsi\mskip 2mu}}}\xspace}
\def\psitwos  {{\ensuremath{\Ppsi{(2S)}}}\xspace}

\def\etac     {{\ensuremath{\Peta_\cquark}}\xspace}
\def\chiczero {{\ensuremath{\Pchi_{\cquark 0}}}\xspace}
\def\chicone  {{\ensuremath{\Pchi_{\cquark 1}}}\xspace}
\def\chictwo  {{\ensuremath{\Pchi_{\cquark 2}}}\xspace}
  \def\Y#1S{\ensuremath{\PUpsilon{(#1S)}}\xspace}

\def\FourS {{\Y4S}}
\def\FiveS {{\Y5S}}

\def\chic  {{\ensuremath{\Pchi_{c}}}\xspace}

\def\proton      {{\ensuremath{\Pp}}\xspace}
\def\antiproton  {{\ensuremath{\overline \proton}}\xspace}

\def\Lz          {{\ensuremath{\PLambda}}\xspace}
\def\Lbar        {{\ensuremath{\kern 0.1em\overline{\kern -0.1em\PLambda}}}\xspace}
\def\LorLbar    {\kern 0.18em\optbar{\kern -0.18em \PLambda}{}\xspace}

\def\Lb      {{\ensuremath{\Lz^0_\bquark}}\xspace}

\def\BF         {{\ensuremath{\mathcal{B}}}\xspace}

\def\BR         {\BF}
\newcommand{\decay}[2]{\ensuremath{#1\!\to #2}\xspace}         
\def\ra                 {\ensuremath{\rightarrow}\xspace}
\def\to                 {\ensuremath{\rightarrow}\xspace}

\def\CP                {{\ensuremath{C\!P}}\xspace}

\def\AT#1     {\ensuremath{A_{\mathrm{T}}^{#1}}\xspace}           

\def\C#1      {\ensuremath{\mathcal{C}_{#1}}\xspace}                       
\def\Cp#1     {\ensuremath{\mathcal{C}_{#1}^{'}}\xspace}                    
\def\Ceff#1   {\ensuremath{\mathcal{C}_{#1}^{\mathrm{(eff)}}}\xspace}        
\def\Cpeff#1  {\ensuremath{\mathcal{C}_{#1}^{'\mathrm{(eff)}}}\xspace}       
\def\Ope#1    {\ensuremath{\mathcal{O}_{#1}}\xspace}                       
\def\Opep#1   {\ensuremath{\mathcal{O}_{#1}^{'}}\xspace}                    

\newcommand{\tev}{\ifthenelse{\boolean{inbibliography}}{\ensuremath{~T\kern -0.05em eV}\xspace}{\ensuremath{\mathrm{\,Te\kern -0.1em V}}}\xspace}
\newcommand{\gev}{\ensuremath{\mathrm{\,Ge\kern -0.1em V}}\xspace}
\newcommand{\mev}{\ensuremath{\mathrm{\,Me\kern -0.1em V}}\xspace}
\newcommand{\kev}{\ensuremath{\mathrm{\,ke\kern -0.1em V}}\xspace}
\newcommand{\ev}{\ensuremath{\mathrm{\,e\kern -0.1em V}}\xspace}
\newcommand{\gevc}{\ensuremath{{\mathrm{\,Ge\kern -0.1em V\!/}c}}\xspace}
\newcommand{\mevc}{\ensuremath{{\mathrm{\,Me\kern -0.1em V\!/}c}}\xspace}
\newcommand{\gevcc}{\ensuremath{{\mathrm{\,Ge\kern -0.1em V\!/}c^2}}\xspace}
\newcommand{\gevgevcccc}{\ensuremath{{\mathrm{\,Ge\kern -0.1em V^2\!/}c^4}}\xspace}
\newcommand{\mevcc}{\ensuremath{{\mathrm{\,Me\kern -0.1em V\!/}c^2}}\xspace}

\def\mum  {\ensuremath{{\,\upmu\mathrm{m}}}\xspace}

\def\invfb   {\ensuremath{\mbox{\,fb}^{-1}}\xspace}

\newcommand{\chisq}{\ensuremath{\chi^2}\xspace}

\def\gsim{{~\raise.15em\hbox{$>$}\kern-.85em
          \lower.35em\hbox{$\sim$}~}\xspace}
\def\lsim{{~\raise.15em\hbox{$<$}\kern-.85em
          \lower.35em\hbox{$\sim$}~}\xspace}

\def\sPlot{\mbox{\em sPlot}\xspace}

\def\sqs   {\ensuremath{\protect\sqrt{s}}\xspace}

\def\ptot       {\mbox{$p$}\xspace}
\def\pt         {\mbox{$p_{\mathrm{ T}}$}\xspace}

\def\evtgen     {\mbox{\textsc{EvtGen}}\xspace}

\def\geant      {\mbox{\textsc{Geant4}}\xspace}

\def\photos     {\mbox{\textsc{Photos}}\xspace}

\def\pythia     {\mbox{\textsc{Pythia}}\xspace}

\def\tell1  {TELL1\xspace}
\def\ukl1   {UKL1\xspace}

\newcommand{\ie}{\mbox{\itshape i.e.}\xspace}

\usepackage{cite} 
\usepackage{mciteplus}

\usepackage{longtable} 

\allowdisplaybreaks

\begin{document}

\renewcommand{\thefootnote}{\fnsymbol{footnote}}
\setcounter{footnote}{1}


\begin{titlepage}
\pagenumbering{roman}

\vspace*{-1.5cm}
\centerline{\large EUROPEAN ORGANIZATION FOR NUCLEAR RESEARCH (CERN)}
\vspace*{0.5cm}
\noindent
\begin{tabular*}{\linewidth}{lc@{\extracolsep{\fill}}r@{\extracolsep{0pt}}}
\ifthenelse{\boolean{pdflatex}}
{\vspace*{-2.7cm}\mbox{\!\!\!\includegraphics[width=.14\textwidth]{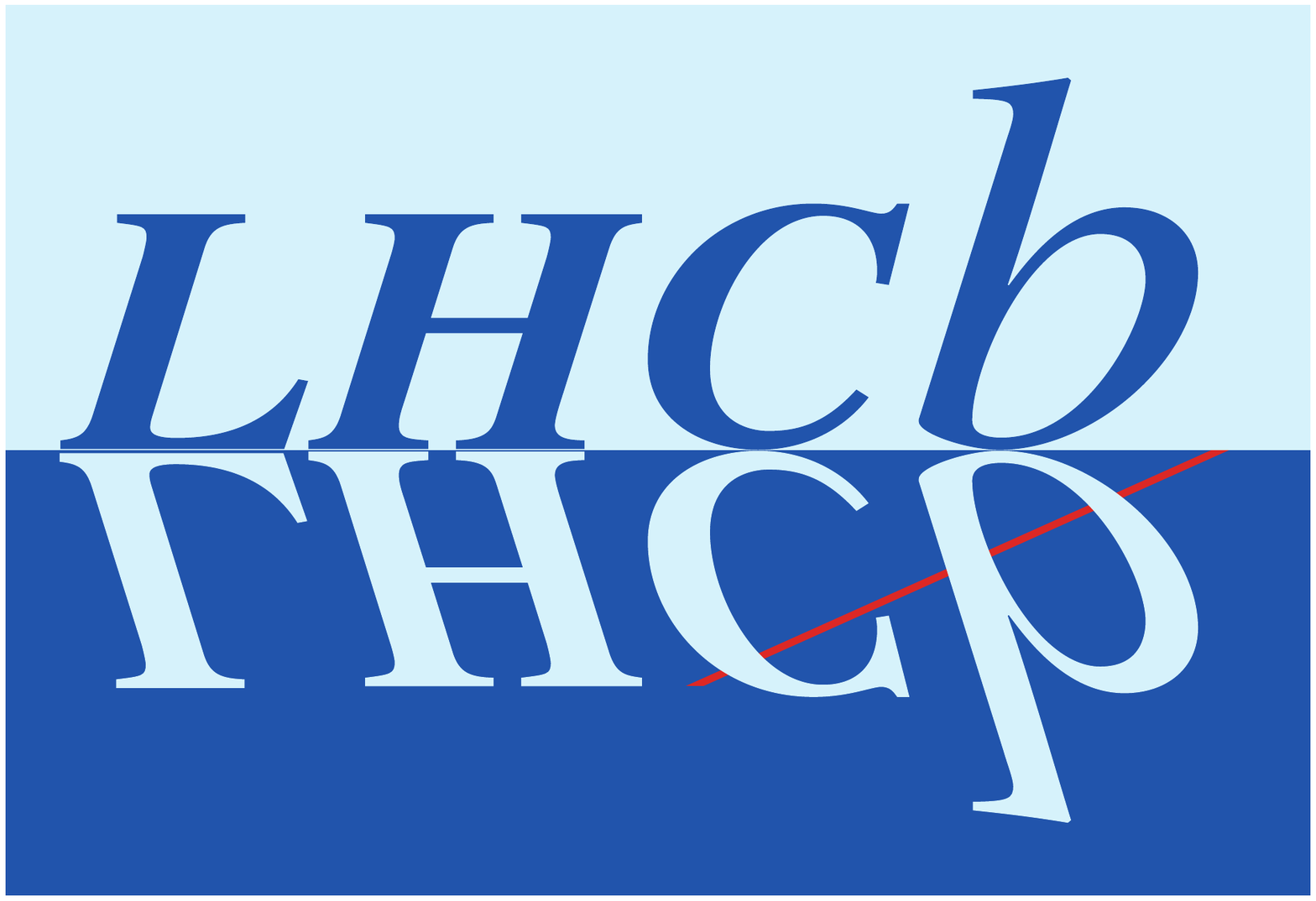}} & &}%
{\vspace*{-1.2cm}\mbox{\!\!\!\includegraphics[width=.12\textwidth]{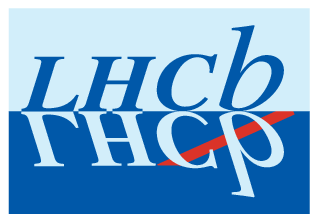}} & &}%
\\
 & & CERN-EP-2017-099 \\  
 & & LHCb-PAPER-2017-007 \\  
 & & \today \\ 
 & & \\
\end{tabular*}

\vspace*{1.0cm}

{\normalfont\bfseries\boldmath\huge
\begin{center}
Study of charmonium production in \bquark-hadron decays and first evidence for the decay $\decay{\Bs}{\phi \phi \phi}$ 
\end{center}
}

\vspace*{0.5cm}

\begin{center}
The LHCb collaboration\footnote{Authors are listed at the end of this paper.}
\end{center}

\vspace{\fill}


\begin{abstract}
\noindent
Using decays to $\phi$-meson pairs, the inclusive production of charmonium states in 
\bquark-hadron decays is studied with $pp$ collision data corresponding to an 
integrated luminosity of $3.0 \invfb$, collected by the LHCb experiment at centre-of-mass energies of 7 and 8 TeV. 
Denoting by $\BR_C \equiv \BR ( \decay{\bquark}{C X} ) \times \BR ( \decay{C}{\phi \phi} )$
the inclusive branching fraction of a $\bquark$ hadron to a charmonium state $C$ that decays into a pair 
of $\phi$ mesons, ratios $R^{C_1}_{C_2}\equiv \BR_{C_1} / \BR_{C_2}$ are determined as
\mbox{$R^{\chiczero}_{\etac (1S)} = 0.147 \pm 0.023 \pm 0.011$}, 
\mbox{$R^{\chicone}_{\etac (1S)} = 0.073 \pm 0.016 \pm 0.006$},
\mbox{$R^{\chictwo}_{\etac (1S)} = 0.081 \pm 0.013 \pm 0.005$},
\mbox{$R^{\chicone}_{\chiczero} = 0.50 \pm 0.11 \pm 0.01$},
$R^{\chictwo}_{\chiczero} = 0.56 \pm 0.10 \pm 0.01$ and
\mbox{$R^{\etac (2S)}_{\etac (1S)} = 0.040 \pm 0.011 \pm 0.004$}.  
Here and below the first uncertainties are statistical and the second 
systematic. Upper limits at 90\% confidence level for the inclusive production of $X(3872)$, 
$X(3915)$ and $\chictwo (2P)$ states are obtained as 
$R^{X(3872)}_{\chicone} < 0.34$, $R^{X(3915)}_{\chiczero} < 0.12$ and 
$R^{\chictwo (2P)}_{\chictwo} < 0.16$.
Differential cross-sections as a function of transverse momentum are measured 
for the $\etac (1S)$ and $\chi_c$ states. 
The branching fraction of the decay 
$\decay{\Bs}{\phi \phi \phi}$ is measured for the first time, 
$\BR ( \decay{\Bs}{\phi \phi \phi} ) = (2.15 \pm 0.54 \pm 0.28 \pm 0.21_{\BR}) \times 10^{-6}$. 
Here the third uncertainty is due to the branching fraction of the decay $\decay{\Bs}{\phi \phi}$, which is used for normalization. 
No evidence for intermediate resonances is seen. 
A preferentially transverse $\phi$ polarization is observed. 
The measurements 
allow the determination of the ratio of the branching fractions for the $\etac (1S)$ decays to $\phi \phi$ 
and $\proton \antiproton$ as 
$\BR ( \decay{\etac (1S)}{\phi \phi} )/\BR ( \decay{\etac (1S)}{\proton \antiproton} ) = 1.79 \pm 0.14\pm 0.32$.
\end{abstract}

\vspace*{0.5cm}

\begin{center}
  Published in Eur.~Phys.~J.~C77 (2017) 609
\end{center}

\vspace{\fill}

{\footnotesize 
\centerline{\copyright~CERN on behalf of the \lhcb collaboration, licence \href{http://creativecommons.org/licenses/by/4.0/}{CC-BY-4.0}.}}
\vspace*{2mm}

\end{titlepage}


\newpage
\setcounter{page}{2}
\mbox{~}
%
%
%
%

\cleardoublepage


\renewcommand{\thefootnote}{\arabic{footnote}}
\setcounter{footnote}{0}



\pagestyle{plain} 
\setcounter{page}{1}
\pagenumbering{arabic}


%

\section{Introduction}
\label{sec:intro}
 
The production of the $J^{PC} = 1^{--}$ charmonium states has been extensively studied using decays to clean dilepton final states. 
Other states such as those from the $\chi_c$ family can be accessed via the radiative transition to \jpsi. 
Studies of the production of the non-$1^{--}$ charmonium states can be performed by reconstructing their decays to fully hadronic final states~\cite{Barsuk:2012ic}. 
This paper reports a measurement of the inclusive production rates of the \etac and $\chi_c$ states in \bquark-hadron decays, 
\mbox{$\decay{\bquark}{\etac X}$} and \mbox{$\decay{\bquark}{\chi_c X}$}, using charmonia decays to a pair of $\phi$ mesons. 
In addition, the first evidence for the decay $\decay{\Bs}{\phi\phi\phi}$ is reported. 

Results on inclusive charmonium production in \bquark-hadron decays are available from \epem experiments operating 
at centre-of-mass energies around the $\FourS$ and $\FiveS$ resonances, studying mixtures of \Bp and \Bz 
mesons\footnote{The inclusion of charge-conjugate states is implied throughout.}
(light mixture) or \Bp, \Bz and \Bs mesons, respectively. 
Mixtures of all \bquark-hadrons (\Bp, \Bz, \Bs, \Bcp and \bquark-baryons)
have been studied at LEP, the Tevatron and the \lhc.
The world average values for charmonium branching fractions from the light mixture are dominated by results from the 
\cleo~\cite{Anderson:2002md,Chen:2000ri}, \belle~\cite{Bhardwaj:2015rju} and \babar~\cite{Aubert:2002hc} collaborations. 
For the \jpsi, \psitwos and \chicone states the measured branching fractions are consistent within uncertainties. 
The new \belle result for the $\decay{\bquark}{\chictwo X}$ branching fraction~\cite{Bhardwaj:2015rju}, 
which supersedes the previous measurement~\cite{Abe:2002wp}, is below the  
\babar result~\cite{Aubert:2002hc} by more than 2.5 standard deviations, while 
the \cleo collaboration does not observe a statistically significant 
$\decay{\bquark}{\chictwo X}$ signal~\cite{Chen:2000ri}. 
An upper limit on the inclusive production rate of $\etac (1S)$ mesons in the light mixture,
$\BF ( \B \to \etac (1S) X ) < 9 \times 10^{-3}$ at $90 \%$~confidence level (CL), was reported by \cleo~\cite{Balest:1994jf}. 

The branching fractions of \bquark-hadron decays to final states including a \jpsi or \psitwos charmonium state, 
where all \bquark-hadron species are involved, are known with uncertainties of around 10\%, with the world averages dominated 
by the measurements performed at \lep~\cite{Abreu:1994rk,Adriani:1993ta,Buskulic:1992wp}. 
The ratio of \mbox{$\decay{\bquark}{\psitwos X}$} and 
\mbox{$\decay{\bquark}{\jpsi X}$} yields has been measured at the \lhc by the \lhcb, \cms and \atlas collaborations 
with a precision of around 5\%~\cite{LHCb-PAPER-2011-045,Chatrchyan:2011kc,Aad:2015duc}. 
The only available results for the \chic family are the \chicone inclusive production rates in 
\bquark-hadron decays measured by the \delphi and \lthree collaborations~\cite{Abreu:1994rk,Adriani:1993ta}, 
with an average value of \mbox{$\BR ( \bquark \to \chicone X) = (14 \pm 4) \times 10^{-3}$}~\cite{PDG2016}. 
Recently, \lhcb measured the $\etac (1S)$ production rate, 
\mbox{$\BR ( \bquark \to \etac (1S) X) = (4.88 \pm 0.64 \pm 0.29 \pm 0.67_{\BR}) \times 10^{-3}$}, 
where the third uncertainty is due to uncertainties on the \jpsi inclusive branching fraction from \bquark-hadron decays 
and the branching fractions of the decays of \jpsi and $\etac (1S)$ to the \proton\antiproton final state~\cite{LHCb-PAPER-2014-029}.

While experimentally the reconstruction of charmonia from \bquark-hadron decays allows an efficient control of combinatorial background 
with respect to charmonium candidates produced in the \proton\proton collision vertex via hadroproduction or in the decays 
of heavier resonances (prompt charmonium), inclusive \bquark-hadron decays to charmonia are theoretically less clean. 
Since a description of the strong interaction dynamics in \bquark-hadron inclusive decays improves with respect to exclusive decays 
due to consideration of more final states, and the formation of charmonium proceeds through a short-distance process,  
a factorization of a $c\bar{c}$ pair production and 
its hadronization in a given charmonium state becomes 
a reasonable assumption~\cite{Bodwin:1994jh}. 
The relative inclusive production of \chic states in \bquark-hadron decays provides a clean test of charmonia production models.
For example, the colour evaporation model predicts a $\chictwo / \chicone$ production ratio of $5 / 3$~\cite{Schuler:1998az},
while the perturbative QCD-based computation predicts that the V-A current, which is responsible for the \bquark decays,
forbids the \chictwo and \chiczero production at leading order. 
In the non-relativistic QCD (NRQCD) framework~\cite{Beneke:1998ks,Beneke:1999gq,Burns:2011rn},
the colour-octet contributions have to be included, predicting the rates to be proportional to $(2J + 1)$ for the $\chic_J$ states. 
The NRQCD framework can be applied to both prompt charmonium production and secondary production from \bquark-hadron decays
and the comparison between these two production mechanisms can provide a valuable test of this theoretical framework.

In this paper we report the first measurements of inclusive $\chi_c$ and $\etac (2S)$ production rates in \bquark-hadron decays 
using charmonium decays to hadronic final states in the high-multiplicity environment of a hadron collider. 
Experimentally, charmonium candidates from \bquark-hadron decays are distinguished from prompt charmonia 
by exploiting the \bquark-hadron decay time and reconstructing a \bquark-hadron (and charmonium) decay vertex 
well separated from the primary vertex where the \bquark-hadron candidate was produced. 
The charmonium states are reconstructed via their decays to a $\phi \phi$ final state. 
The $\etac (1S)$ production followed by the decay $\etac (1S) \to \phi \phi$ is used for normalization, so that systematic uncertainties partially cancel in the ratios. 
As a by-product of the production rate measurements, the masses of the $\etac (1S)$, \chiczero, \chicone, \chictwo and $\etac (2S)$ 
charmonium states and the natural width of the $\etac (1S)$ meson are determined. 

The \Bs decay to the $\phi \phi$ final state has been observed by the \cdf collaboration~\cite{Aaltonen:2011rs} 
and recently precisely measured by the \lhcb collaboration~\cite{LHCb-PAPER-2015-028}, 
where it was also used to search for \CP-violating asymmetries~\cite{LHCb-PAPER-2014-026}. 
In the Standard Model (SM) the amplitude for the decay $\Bs \ra \phi \phi$ is dominated by a loop diagram. 
Experimental verification of the partial width, polarization amplitudes and triple-product asymmetries of the $\Bs \to \phi \phi$ decay 
probes the QCD contribution to the weak processes described by 
nonfactorizable penguin diagrams~\cite{Kagan:2004uw,Datta:2007qb}, 
and contributions from particles beyond the SM 
to the penguin loops~\cite{Chen:2005mka,Huang:2005qb,Bartsch:2008ps,Beneke:2006hg,Cheng:2009mu}. 
A three-body $\Bs \to \phi \phi \phi$ decay leads to a final state with six strange quarks. 
In the SM it is described by the penguin diagram 
of the $\Bs \to \phi \phi$ decay with the creation of an additional \ssbar quark pair. 
The $\Bs \to \phi \phi \phi$ decay can also receive contributions from an intermediate 
charmonium state decaying to a $\phi \phi$ state. 
Here we report first evidence for the $\Bs \to \phi \phi \phi$ decay and study its resonance structure. 
The branching fraction of this decay is determined relative to the branching fraction $\BR ( \Bs \to \phi \phi )$~\cite{LHCb-PAPER-2015-028}. 
To cross-check the technique exploited in this paper, the value of $\BR ( \Bs \to \phi \phi )$ 
is also determined relative to the $\etac (1S)$ production rate. 
Finally, the ratio of the branching fractions for the decays $\etac (1S) \to \phi \phi$ and \mbox{$\etac (1S) \to \proton \antiproton$} 
is determined, using additional external information. 

The \lhcb detector and data sample used for the analysis are presented in Sect.~\ref{sec:lhcb}.
Section~\ref{sec:sel} explains the selection details and the signal extraction technique. 
Inclusive production of charmonium states in \bquark-hadron decays is discussed in Sect.~\ref{sec:ccsection}.
In Sect.~\ref{sec:masses} measurements of the $\etac (1S)$ mass and natural width are described. 
First evidence for the $\Bs \to \phi \phi \phi$ decay is reported in Sect.~\ref{sec:bs}. 
The main results of the paper are summarized in Sect.~\ref{sec:summary}.

\section{\lhcb detector and data sample}
\label{sec:lhcb}

The \lhcb detector~\cite{Alves:2008zz,LHCb-DP-2014-002} is a single-arm forward spectrometer covering 
the \mbox{pseudorapidity} range $2<\eta <5$, designed for the study of particles containing \bquark or \cquark quarks. 
The detector includes a high-precision tracking system consisting of a silicon-strip vertex detector surrounding the $pp$
interaction region, a large-area silicon-strip detector located upstream of a dipole magnet with a bending power of about
$4{\mathrm{\,Tm}}$, and three stations of silicon-strip detectors and straw drift tubes placed downstream of the magnet.
The tracking system provides a measurement of momentum, \ptot, of charged particles with a relative uncertainty that varies 
from 0.5\% at low momentum\footnote{Natural units are used throughout the paper.} to 1.0\% at 200\gev.
The minimum distance of a track to a primary vertex (PV), the impact parameter (IP), is measured with a resolution of $(15+29/\pt)\mum$,
where \pt is the component of the momentum transverse to the beam, in\,\gev.
Different types of charged hadrons are distinguished using information from two ring-imaging Cherenkov detectors. 
Photons, electrons and hadrons are identified by a calorimeter system consisting of scintillating-pad and preshower detectors, 
an electromagnetic calorimeter and a hadronic calorimeter. 
Muons are identified by a system composed of alternating layers of iron and multiwire proportional chambers. 
The online event selection is performed by a trigger, 
which consists of a hardware stage, based on information from the calorimeter and muon
systems, followed by a software stage, which applies a full event reconstruction.

The analysis is based on \proton\proton collision data recorded by the \lhcb experiment 
at a centre-of-mass energy $\sqs = 7 \tev$, corresponding to an integrated luminosity of $1.0 \invfb$, 
and at $\sqs = 8 \tev$, corresponding to an integrated luminosity of $2.0 \invfb$.
Events enriched in signal decays are selected by the hardware trigger, based on the presence of a single deposit of high transverse energy 
in the calorimeter.
The subsequent software trigger selects events with displaced vertices formed by 
charged particles having a good track-fit quality, transverse momentum larger than 0.5\gev, 
and that are incompatible with originating from any PV~\cite{LHCb-PAPER-2014-026}. 
Charged kaon candidates are identified using the information from the Cherenkov and tracking detectors. 
Two oppositely charged kaon candidates having an invariant mass within $\pm 11 \mev$ of the known mass of the $\phi$ meson 
are required to form a good quality vertex. 

Precise mass measurements require a momentum-scale calibration. 
The procedure~\cite{LHCb-PAPER-2012-048} uses $\jpsi \to \mup \mun$ decays to cross-calibrate a relative momentum scale 
between different data-taking periods. 
The absolute scale is determined using $\Bp \to \jpsi \Kp$ decays with known particle masses as input~\cite{PDG2016}. 
The final calibration is checked with a variety of fully reconstructed quarkonium, \Bp and \KS decays. 
No residual bias is observed within the experimental resolution.

In the simulation, $pp$ collisions are generated using \pythia~\cite{Sjostrand:2006za,Sjostrand:2007gs} 
with a specific \lhcb configuration~\cite{LHCb-PROC-2010-056}.  
Decays of hadronic particles are described by \evtgen~\cite{Lange:2001uf}, 
in which final-state radiation is generated using \photos~\cite{Golonka:2005pn}. 
The interaction of the generated particles with the detector material and its response are implemented 
using the \geant toolkit~\cite{Allison:2006ve, *Agostinelli:2002hh} as described in Ref.~\cite{LHCb-PROC-2011-006}.
Simulated samples of $\etac$ and $\chi_c$ mesons decaying to the $\phi \phi$ final state, 
and \Bs decaying to two and three $\phi$ mesons, are used to estimate efficiency ratios
and to evaluate systematic uncertainties. 
Charmonium and $\Bs \to \phi \phi \phi$ decays are generated with uniform phase-space density, 
while $\Bs \to \phi \phi$ decays are generated according to the amplitudes from Ref.~\cite{Aaltonen:2011rs}.

\section{Selection criteria and signal extraction} 
\label{sec:sel}

The signal selection is largely performed at the trigger level. 
The offline analysis selects combinations of two or three $\phi$ candidates that are required to form a good-quality common vertex displaced 
from the primary vertex. 
A good separation between the two vertices ($\chisq > 100$) is required, reducing the contribution 
from charmonia produced directly at the primary vertex to below $1 \%$.
Pairs of $\phi$ mesons originating from different \bquark-hadrons produced in the same beam crossing event are suppressed by the requirement of a good-quality 
common vertex. 
Detector acceptance and selection requirements, and in particular the requirement of the kaon \pt to exceed 0.5\gev, 
significantly suppress $\phi \phi$ combinations with \pt below $4\gev$. 

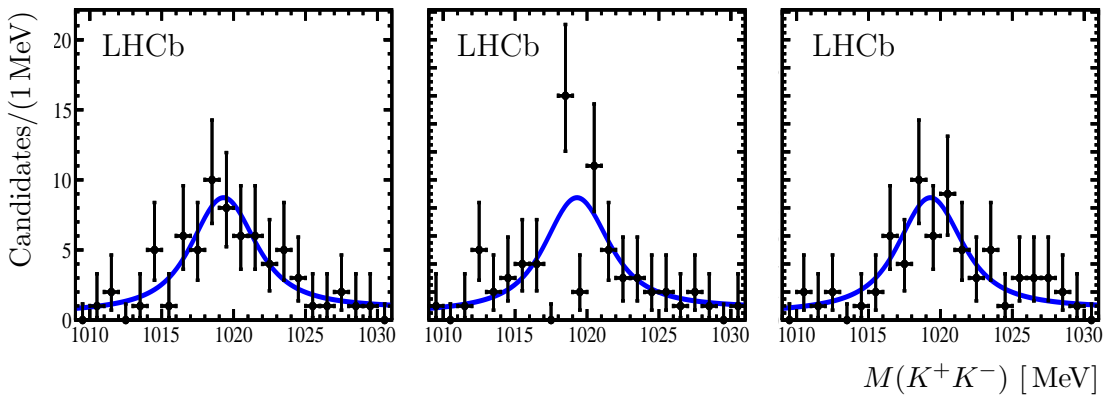
\begin{figure}[t]
\centering
\begin{picture}(450,157)
\put(30,5){\includegraphics[width=405px,height=5.3cm]{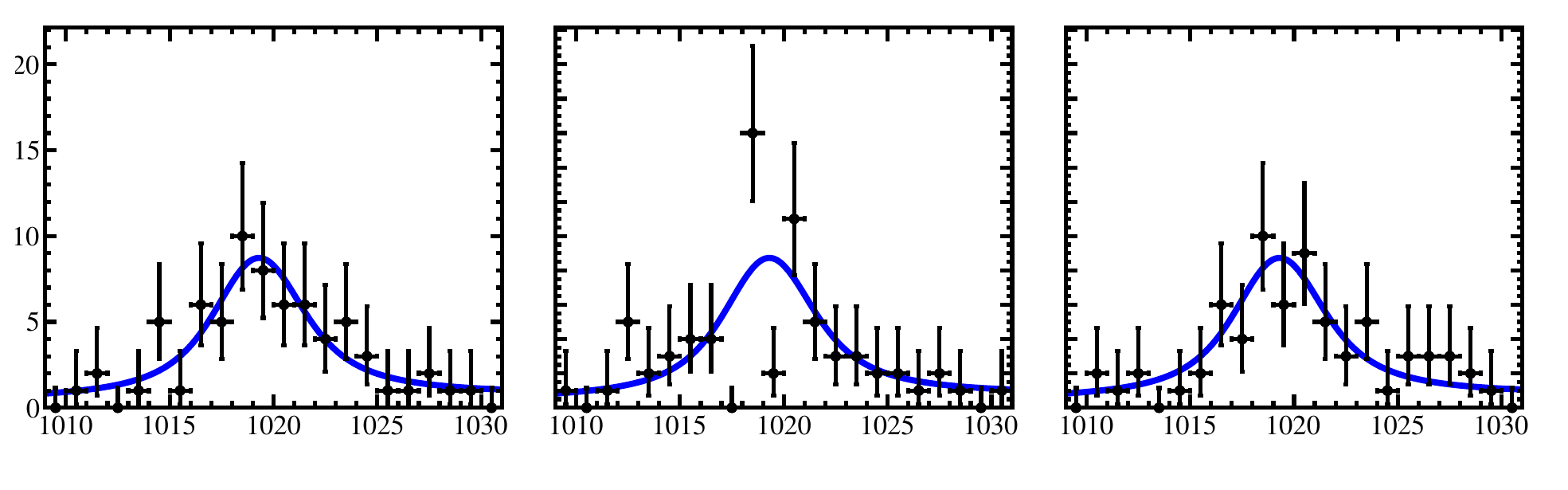}
                \put(-420,45){\rotatebox{90}{\small{Candidates/$( 1 \mev )$}}}
                \put(-385,125){\lhcb}
                \put(-252,125){\lhcb}
                \put(-119,125){\lhcb}
                \put(-100,0) {\small{$M( \Kp \Km ) \ [ \mev ]$}}}
\end{picture}
\protect\caption{Projections to the $\Kp \Km$ invariant mass axes of the 3D fit to the $3 ( K^+ K^- )$ invariant mass distribution 
in the \Bs signal region.}
\label{fig:threedbs}
\end{figure}

Two-dimensional (2D) or three-dimensional (3D) extended unbinned maximum likelihood fits, corresponding to the two or three 
randomly ordered $K^+ K^-$ combinations, 
are performed in bins of the invariant mass of the four-kaon and six-kaon combinations, denoted as $2 ( K^+ K^- )$ or $3 ( K^+ K^- )$, respectively, 
to determine the numbers of $\phi \phi$ and $\phi \phi \phi$ combinations. 
The 2D fit accounts for the signal, $\phi \phi$, and background, $\phi \, ( K^+ K^- )$ and $2 ( K^+ K^- )$, components, 
while the 3D fit accounts for the signal, $\phi \phi \phi$, and background, $\phi \phi \, ( K^+ K^- )$, $\phi \, 2( K^+ K^- )$ and $3 ( K^+ K^- )$, components. 
A $\phi$ signal is described by the convolution of a Breit-Wigner function 
and a sum of two Gaussian functions with a common mean. 
The ratio of the two Gaussian widths and the fraction of the narrower Gaussian are taken from simulation. 
The contribution from combinatorial background, due to $\Kp \Km$ pairs not originating from the decay of a $\phi$ meson, is assumed to be flat. 
In addition, a threshold function to account for the available phase-space in the $\Kp \Km$ system is introduced for both signal and combinatorial background.
While no visible contribution from the $f_0 (980)$ resonance decaying into a $K^+ K^-$ pair is observed  
in the $2 ( K^+ K^- )$ or $3 ( K^+ K^- )$ combinations, a potential effect due to contributions from such decays is considered as a source of systematic uncertainty. 
\begin{figure}[b]
\centering
\begin{picture}(450,150)
\put(15,6){\includegraphics[width=450px]{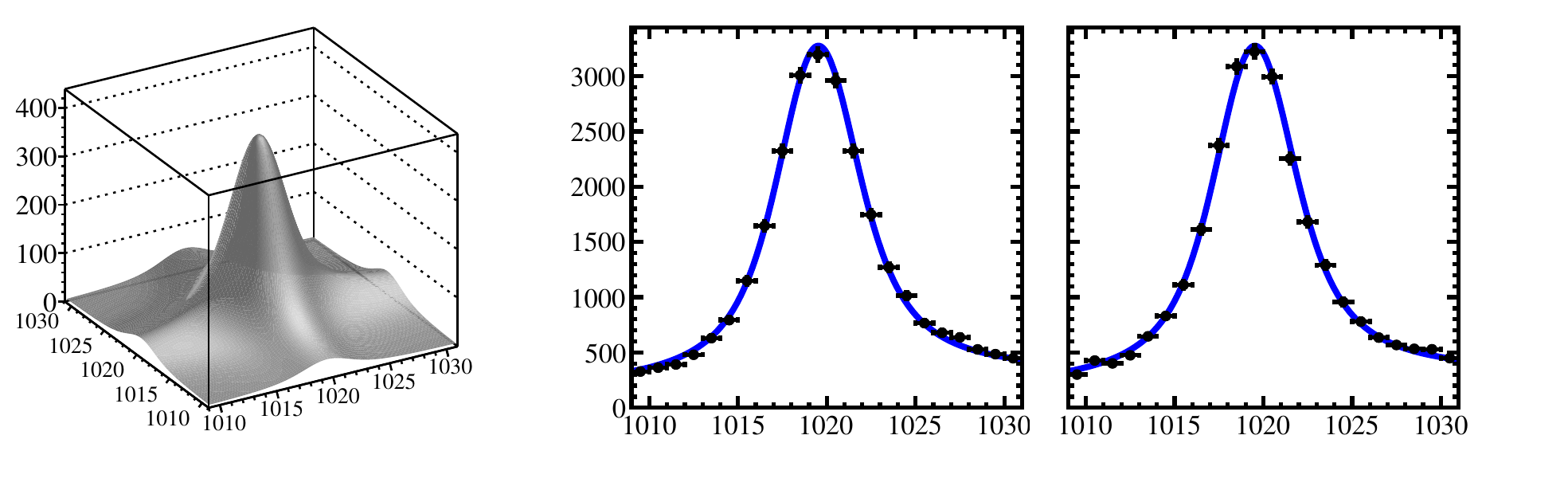}
                \put(-300,35){\rotatebox{90}{\small{Candidates/$( 1 \mev )$}}}
                \put(-458,50){\rotatebox{90}{\scriptsize{Candidates/$( 0.22 \mev )^2$}}}
                \put(-265,115){\lhcb}
                \put(-140,115){\lhcb}
                \put(-120,0) {\small{$M( \Kp \Km ) \ [ \mev ]$}}
                \put(-385,8) {\rotatebox{13}{\scriptsize{$M( \Kp \Km ) \ [ \mev ]$}}}
                \put(-455,40) {\rotatebox{322}{\scriptsize{$M( \Kp \Km ) \ [ \mev ]$}}}}
\end{picture}
\protect\caption{Result of the 2D fit to the $2 ( K^+ K^- )$ invariant mass distribution 
along with the projections to the $\Kp \Km$ invariant mass axes in the $\etac (1S)$ signal region.} 
\label{fig:twodetac}
\end{figure}
\begin{figure}[h]
\centering
\begin{picture}(450,148)
\put(15,6){\includegraphics[width=450px]{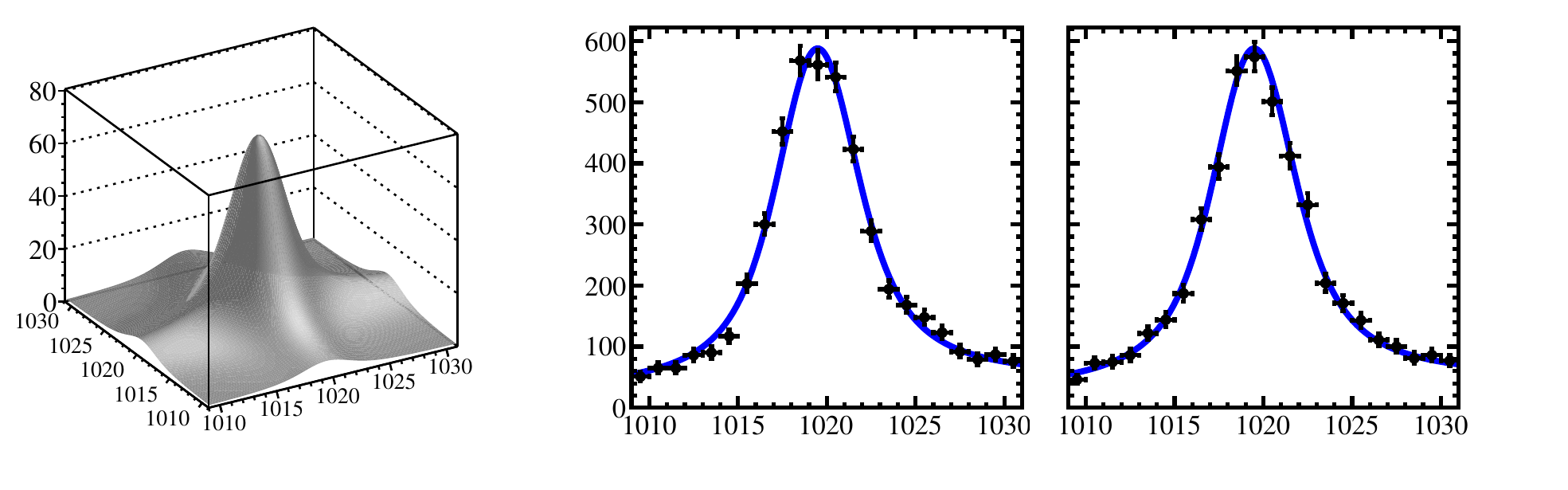}
                \put(-297,35){\rotatebox{90}{\small{Candidates/$( 1 \mev )$}}}
                \put(-455,50){\rotatebox{90}{\scriptsize{Candidates/$( 0.22 \mev )^2$}}}
                \put(-265,115){\lhcb}
                \put(-140,115){\lhcb}
                \put(-120,0) {\small{$M( \Kp \Km ) \ [ \mev ]$}}
                \put(-385,8) {\rotatebox{13}{\scriptsize{$M( \Kp \Km ) \ [ \mev ]$}}}
                \put(-455,40) {\rotatebox{322}{\scriptsize{$M( \Kp \Km ) \ [ \mev ]$}}}}
\end{picture}
\protect\caption{Result of the 2D fit to the $2 ( K^+ K^- )$ invariant mass distribution 
along with the projections to the $\Kp \Km$ invariant mass axes in the \Bs signal region.} 
\label{fig:twodbs}
\end{figure}
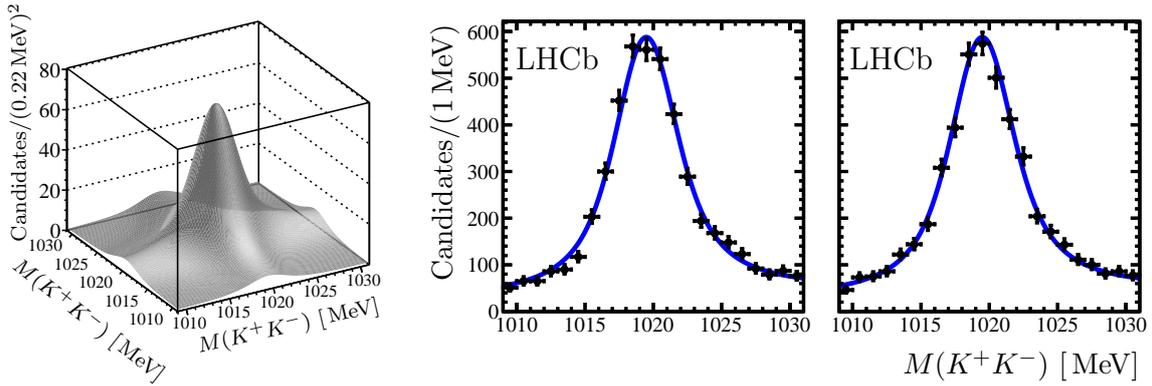
Figures~\ref{fig:twodetac} and \ref{fig:twodbs} show the results of the 2D fits to the $2 ( K^+ K^- )$ invariant mass 
distributions
along with the projections to the $\Kp \Km$ invariant mass axes in the $\etac (1S)$ and \Bs signal regions, 
$2.91 - 3.06 \gev$ and $5.30 - 5.43 \gev$.  
Figure~\ref{fig:threedbs} shows the projections to the $\Kp \Km$ invariant mass axes of the 3D fit to the $3 ( K^+ K^- )$ 
invariant mass distribution in the \Bs signal region. 
While using the known value for the natural width of the $\phi$ resonance~\cite{PDG2016}, 
the $\phi$ mass and the remaining resolution parameter are determined from the fits in the enlarged signal $\phi\phi$ and 
$\phi\phi\phi$ invariant mass regions. 
In the 2D and 3D fits in the bins of $\phi\phi$ and $\phi\phi\phi$ invariant mass
the $\phi$ mass and the resolution parameter are then fixed to 
the values determined in the enlarged signal regions. 

Unless they are extracted from the 2D or 3D fits, throughout the paper
the error bars shown in the histograms are determined as follows: the upper (lower) error bar assigned 
to a given bin content $N$ is defined by the expectation value $\lambda$ of the Poissonian distribution giving 16\% probability 
to observe the number of events $N$ or less (more). 
When obtained from the 2D or 3D fits the histogram bin contents are constrained to be positive, 
with error bars defined by the range in the allowed region where the best fit negative-log likelihood value is within 
half a unit from the global minimum.

In the following, production ratios are determined from the signal yields obtained from the fits 
of the $\phi \phi$ or $\phi \phi \phi$ invariant mass spectra. 
The relative efficiencies are taken into account to determine the ratio of the branching fractions of the corresponding decays. 
Signal yields corresponding to the data samples accumulated at $\sqs = 7$ and $8 \tev$ 
are found to be compatible. 
Unless otherwise specified, the results below are based on the analysis of the combined data sample.

\protect\section{Charmonium production in decays to $\boldsymbol{\phi \phi}$}
\label{sec:ccsection}

\subsection{Charmonium yields}

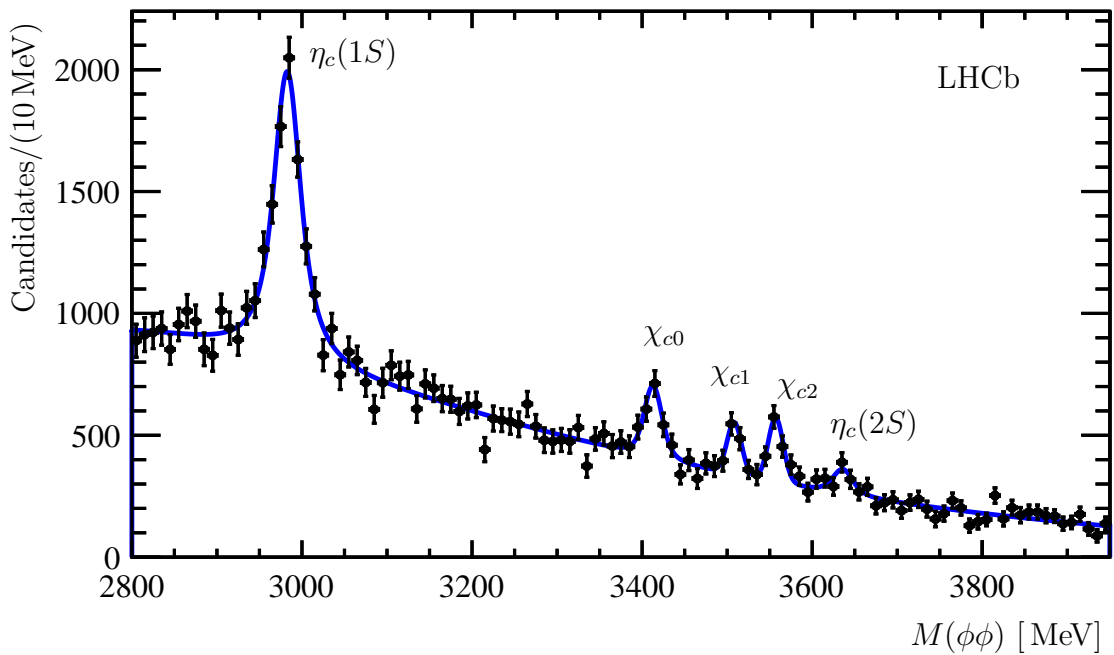
\begin{figure}[b]
\centering
\begin{picture}(430,260)
\put(0,10){\includegraphics[width=450px]{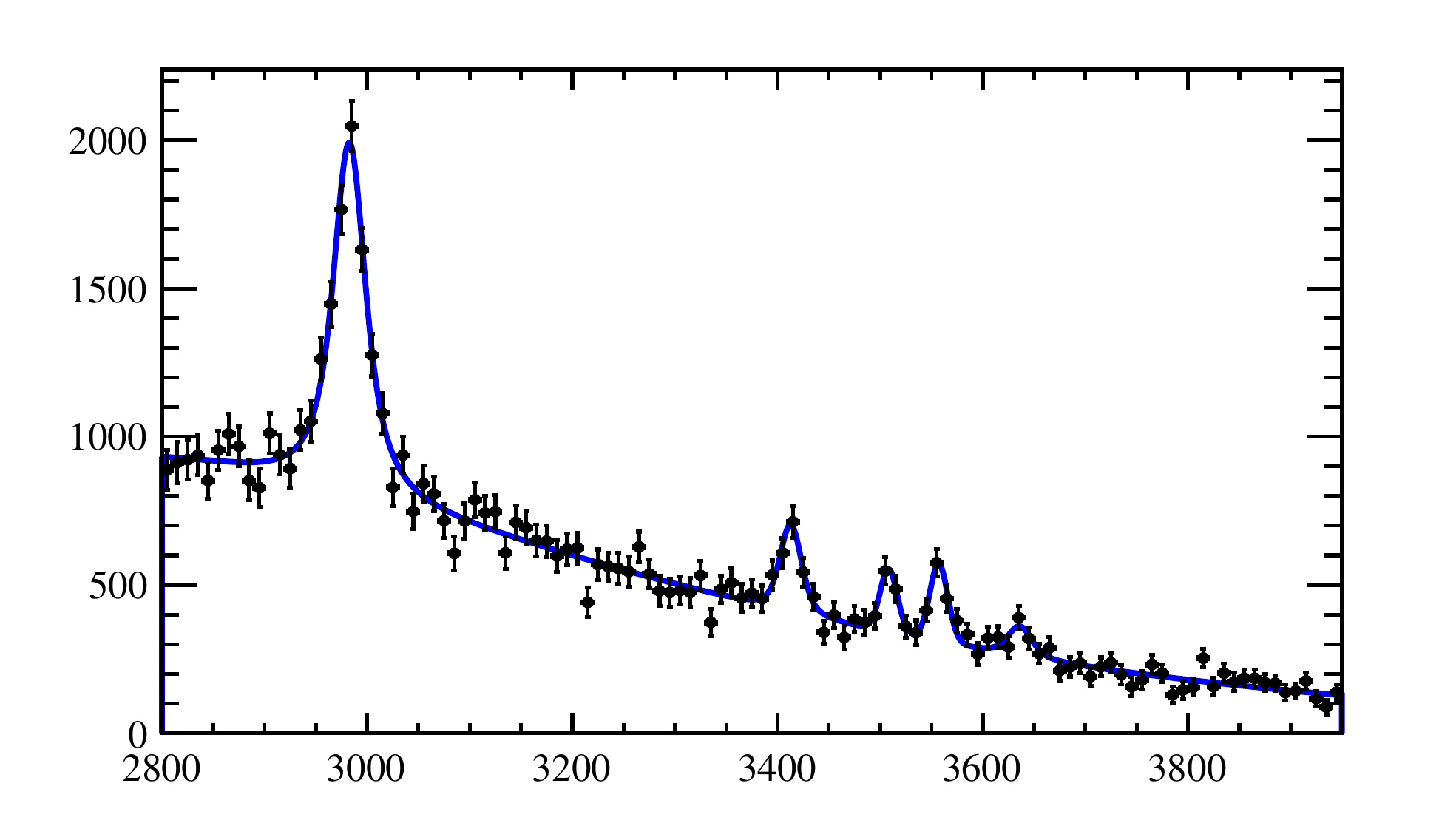}
                \put(-447,125){\rotatebox{90}{{Candidates/$( 10 \mev )$}}}
                \put(-100,210) {\lhcb}
                \put(-335,220) {$\etac(1S)$}
                \put(-210,115) {\chiczero}
                \put(-185,100) {\chicone}
                \put(-160,95) {\chictwo}
                \put(-140,80) {$\etac(2S)$}
                \put(-110,0) {{$M( \phi \phi ) \ [ \mev ]$}}}
\end{picture}
\caption{Distribution of the invariant mass of $\phi \phi$ combinations. 
The number of candidates in each bin is obtained from the corresponding 2D fit. 
The peaks corresponding to the \ccbar resonances are marked on the plot. 
The signal yields are given in Table~\ref{tab:cc}.} \label{fig:ccphiphi}
\end{figure}

Figure~\ref{fig:ccphiphi} shows the invariant mass spectrum of the $\phi \phi$ combinations,
where the content of each bin is a result of a 2D fit to the two $\Kp \Km$ invariant-mass combinations. 
A binned $\chi^2$ fit to the spectrum is used to determine 
the contributions from the $\etac (1S)$ and $\etac (2S)$ mesons, and the \chiczero, \chicone and \chictwo mesons. 
The charmonium-like states $X(3872)$, $X(3915)$ and $\chictwo (2P)$ with masses and natural widths from Ref.~\cite{PDG2016} 
are taken into account in alternative fits 
in order to evaluate systematic uncertainties, as well as to obtain upper limits on the inclusive production of these states in \bquark-hadron decays. 
Each resonance is described by the convolution of a relativistic Breit-Wigner function and a sum of two Gaussian functions with a common mean. 
The combinatorial background, \ie contributions due to random combinations of two true $\phi$ mesons, 
is described by the product of a first-order polynomial with an exponential function and a threshold factor. 
The natural width of the $\etac (1S)$ state is a free parameter in the fit, 
while the natural widths of the $\etac (2S)$ and the $\chi_c$ states, which have lower signal yields, 
are fixed to their known values~\cite{PDG2016}. 
Possible variations of the $\etac (2S)$ production rate depending on its natural width, $\Gamma_{\etac (2S)}$, are explored 
by providing the result as a function of the assumed natural width. 
The ratio of the two Gaussian widths and the fraction of the narrow Gaussian are fixed from the simulation. 
The mass resolution for different charmonium resonances 
is scaled according to the energy release, so that a single free parameter in the $\phi \phi$ invariant mass fit accounts for the detector resolution. 
This scaling of the mass resolution for different charmonium states has been validated using simulation.

\addtolength{\tabcolsep}{-4pt}
\begin{table}[t]
\centering
\caption[Charmonium signal yields from the fit to $\phi \phi$ invariant mass spectrum]{Signal yields with statistical uncertainties 
of the fit to the spectrum of the $\phi \phi$ invariant mass. 
\label{tab:cc}}
\begin{tabular}{l|rcl}
 Resonance$\ $ & \multicolumn{3}{c}{$\, $ Signal yield}  \\ 
\hline
$\ \ \ \etac (1S)$   & $\ $ 6476 & $\pm$ & 418  \\ 
$\ \ $ \chiczero   & 933 & $\pm$ & 128   \\
$\ \ $ \chicone    & 460 & $\pm$ & 89   \\ 
$\ \ $ \chictwo    & 611 & $\pm$ & 97  \\
$\ \ \ \etac (2S)$   & 365 & $\pm$ & 100
\end{tabular}
\end{table}
\addtolength{\tabcolsep}{4pt}
\begin{table}[b]
\centering
\caption{The ratio of charmonium signal yields with respect to the $\etac (1S)$ yield and between pairs of \chic states. 
The first uncertainties are statistical and the second systematic.
\label{tab:ccc}}
\begin{tabular}{r|c}
 Resonances & Signal yield ratio \\ 
\hline
$N_{\chiczero} / N_{\etac (1S)}$  & $0.144 \pm 0.022 \pm 0.011$  \\ 
$N_{\chicone} / N_{\etac (1S)}$   & $0.071 \pm 0.015 \pm 0.006$  \\ 
$N_{\chictwo} / N_{\etac (1S)}$   & $0.094 \pm 0.016 \pm 0.006$  \\
$N_{\chicone} / N_{\chiczero}$   & $0.494 \pm 0.107 \pm 0.012$  \\ 
$N_{\chictwo} / N_{\chiczero}$   & $0.656 \pm 0.121 \pm 0.015$  \\ 
$N_{\etac (2S)} / N_{\etac (1S)}$  & $0.056 \pm 0.016 \pm 0.005$  \\ 
\end{tabular}
\end{table}
The signal yields are given in Table~\ref{tab:cc}.  
The ratios of the resonance yields from the fit are given in Table~\ref{tab:ccc}, both
for the ratios with respect to the $\etac (1S)$ yield and between pairs of \chic states; 
the systematic uncertainties are discussed below.  
The statistical significance for the $N_{\etac (2S)}$ signal is estimated from the \chisq-profile 
to be 3.7 standard deviations. 

Systematic uncertainties in the ratios of the charmonium yields are estimated by considering potential contributions 
from other states, from imperfect modelling of detector resolution, signal resonances and background, and from the 2D fit technique. 
In order to evaluate the systematic uncertainty related to potential contributions
from other states, signal shapes for the $X (3872)$, $X(3915)$, and $\chictwo (2P)$ states are included in the fit. 
Systematic uncertainties related to detector resolution are estimated by fixing 
the $\etac (1S)$ mass resolution to the value determined from the simulation. 
In addition, systematic uncertainties associated to the impact of the detector resolution 
on the signal shapes are estimated by comparing the nominal fit results to those obtained 
using a single instead of a double Gaussian function. 
An uncertainty associated with the description of the mass resolution of the $\phi$ meson is estimated 
by fixing the resolution in the 2D fits to the value determined from simulation. 
The uncertainty associated with the description using the relativistic Breit-Wigner line shape~\cite{Jackson:1964zd} is estimated 
by varying the radial parameter of the Blatt-Weisskopf barrier factor~\cite{Blatt:1952ije} between $0.5$ and $3 \, \gev^{-1}$. 
In order to estimate the uncertainty related to the natural width of the $\etac (2S)$ meson, the value of $\Gamma_{\etac (2S)}$ 
is varied within the uncertainties of the world average~\cite{PDG2016}. 
The uncertainty in the description of the $\chi_c$ signal peaks is estimated by fixing the $\chi_c$ masses to their known values. 
A reduced fit range, covering only the $\chi_c$ and $\etac (2S)$ region ($3.15 - 3.95 \gev$), 
is used to estimate a systematic uncertainty associated to the choice of the fit range. 
An alternative background parametrization using a parabolic instead of a linear function is used 
to estimate the systematic uncertainty due to the choice of the background parametrisation. 
A systematic uncertainty associated to the background parametrization in the 2D fits is estimated 
by adding slope parameters to the description of the non-$\phi$ $\Kp \Km$ combinations 
in the $\phi \Kp \Km$ and the $2 \times (\Kp \Km)$ components. 
The effect of a potential contribution from the $f_0 (980)$ state in the 2D fits
is estimated by including the $f_0 (980)$ contribution 
following the analysis described in Ref.~\cite{LHCb-PAPER-2011-002}, 
and varying the $f_0 (980)$ mass and natural width within the uncertainties quoted in Ref.~\cite{PDG2016}. 
Contributions from multiple candidates are below 2\% per event and are not considered in the evaluation of systematic uncertainties. 
The uncertainty related to the momentum-scale calibration is negligible. 

The total systematic uncertainty is obtained as the quadratic sum of the individual systematic contributions. 
The systematic uncertainties are shown in Table~\ref{tab:cccsyst}.
\begin{table}[t]
\centering
\caption{Systematic uncertainties of the charmonium event yield ratios 
within families and with respect to the $\etac (1S)$ yield. 
The total uncertainty is the sum in quadrature of the individual contributions.
\label{tab:cccsyst}}
\resizebox{\textwidth}{!}{%
\begin{tabular}{l|r|r|r|r|r|r}
Systematic uncertainty & $\frac{N_{\chiczero}}{N_{\etac(1S)}}$ & $\frac{N_{\chicone}}{N_{\etac(1S)}}$ & $\frac{N_{\chictwo}}{N_{\etac(1S)}}$ & 
$\frac{N_{\chicone}}{N_{\chiczero}}$ & $\frac{N_{\chictwo}}{N_{\chiczero}}$ & $\frac{N_{\etac(2S)}}{N_{\etac(1S)}}$ \\ \hline 
Including other states                   & 0.004 & 0.003 & 0.003    & $ 0.006$ & $ 0.008$ & $0.003$ \\
Description of detector resolution       & $< 0.001$ & $< 0.001$ & $< 0.001$ & $0.001$ & $0.001$  & $0.002$ \\
Description of signal resonances         & $0.002$ & $0.002$ & $0.001$ & $0.010$ & $0.002$  & $0.003$ \\
Background model                         & $0.010$ & $0.005$ & $0.005$ & 0.002    & 0.012    &$< 0.001$ \\ 
2D fit functions                         & $ 0.002$ & $ < 0.001$ & $0.001$ & $ 0.005$ & $ 0.005$ & $ 0.001$ \\ \hline
                    Total             & $0.011$   & $0.006$   & $0.006$   & $0.012$   & $0.015$ & $0.005$ \\
\end{tabular}}
\end{table}
The description of the background and the potential contributions from other resonances dominate the total systematic uncertainties. 
In the yield ratios the systematic uncertainty is smaller than or comparable to the statistical uncertainty. 

Complementary cross-checks are performed by comparing the distributions of kinematic variables in simulation and data. 
The stability of the results is checked by using an alternative binning of the $\phi \phi$ invariant mass distribution 
(shifted by half a bin) and by using the weighted signal events from the \sPlot~\cite{Pivk:2004ty} instead of the nominal 2D fit technique. 
The same check is performed for the determination of the masses and widths of the charmonium states. 
In all cases no significant changes are observed and no additional contributions to the systematic uncertainties 
are assigned.

\protect\subsection{Production of $\boldsymbol{\etac}$ and $\boldsymbol{\chi_c}$ in \bquark-hadron decays}

The production ratios of charmonium $C$ with respect to the $\etac (1S)$ yield and between pairs of \chic states
\begin{equation*}
    R^{C_1}_{C_2} \equiv \frac{\BR ( \bquark \to C_1\,X ) \times \BR ( C_1 \to \phi \, \phi )}{\BR ( \bquark \to C_2\,X ) \times \BR ( C_2 \to \phi \, \phi )}
\end{equation*}
are determined to be 
\begin{align*}
R^{\chiczero}_{\etac (1S)} &= 0.147 \pm 0.023 \pm 0.011 , \\
R^{\chicone}_{\etac (1S)} &= 0.073 \pm 0.016 \pm 0.006 , \\
R^{\chictwo}_{\etac (1S)} &= 0.081 \pm 0.013 \pm 0.005 , \\
R^{\chicone}_{\chiczero} &= 0.50 \pm 0.11 \pm 0.01 , \\
R^{\chictwo}_{\chiczero} &= 0.56 \pm 0.10 \pm 0.01 , \\
R^{\etac (2S)}_{\etac (1S)} &= 0.040 \pm 0.011 \pm 0.004 , 
\end{align*}
where, here and throughout, the first uncertainties are statistical and the second are systematic. 
The dominant contributions to the systematic uncertainty on the relative $\chi_c$ production rates 
arise from accounting for possible other resonances and from uncertainties on the $\chi_c$ masses~\cite{PDG2016}. 
The systematic uncertainties are smaller than the statistical uncertainties, 
so that the overall uncertainty on the measurements will be reduced with a larger dataset. 
The systematic uncertainty on the relative production rate of $\etac (2S)$ mesons is dominated by possible contributions from other resonances 
and variation of the $\etac (2S)$ natural width. 

In the following, the $\etac (1S)$ production rate in \bquark-hadron decays and the branching fractions of the charmonium decays to $\phi \phi$ are used 
to obtain single ratios of charmonium production rates in \bquark-hadron decays and the branching fractions of inclusive \bquark-hadron transitions 
to charmonium states. 
The $\etac (1S)$ inclusive production rate in \bquark-hadron decays was measured by \lhcb as 
$\BR ( b \to \etac (1S) X ) = ( 4.88 \pm 0.97 ) \times 10^{-3}$~\cite{LHCb-PAPER-2014-029} using decays to \proton\antiproton. 
Branching fractions of the charmonia decays to $\phi \phi$ from Ref.~\cite{PDG2016} are used. 
However, Ref.~\cite{PDG2016} indicates a possible discrepancy for the $\BR ( \etac (1S) \to \phi \phi )$ value when comparing 
a direct determination and a fit including all available measurements. 
Therefore, an average of the results from \belle~\cite{Huang:2003dr} and \babar~\cite{Aubert:2004gc} using \Bp decays to $\phi \phi \Kp$, 
$\BR ( \etac (1S) \to \phi \phi ) = ( 3.21 \pm 0.72 ) \times 10^{-3}$, is used below. 
The uncertainty of this average dominates the majority of the following results in this section, and an improved knowledge 
of the $\BR ( \etac (1S) \to \phi \phi )$ is critical to reduce the uncertainties of the derived results. 
The values $\BR ( \chiczero \to \phi \phi ) = (7.7 \pm 0.7)  \times 10^{-4}$, $\BR ( \chicone \to \phi \phi ) = (4.2 \pm 0.5)  \times 10^{-4}$, 
and $\BR ( \chictwo \to \phi \phi ) = (1.12 \pm 0.10) \times 10^{-3}$, are used for the $\chi_c$ decays~\cite{PDG2016}. 
The relative branching fractions of \bquark-hadron inclusive decays into $\chi_c$ states are then derived to be 
\begin{align*}
\frac{\BR ( b \to \chicone X )}{\BR ( b \to \chiczero X )}  &= 0.92 \pm 0.20 \pm 0.02 \pm 0.14_{\BR} , \\
\frac{\BR ( b \to \chictwo X )}{\BR ( b \to \chiczero X )}  &= 0.38 \pm 0.07 \pm 0.01 \pm 0.05_{\BR} , 
\end{align*}
where
the third uncertainty is due to those on the branching fractions $\BR ( \chi_c \to \phi \phi )$. 
The result for the relative \chicone and \chictwo production in inclusive \bquark-hadron decays is close to the values measured 
in \Bz and \Bp production~\cite{PDG2016}. 

The branching fractions of \bquark-hadron decays into $\chi_c$ states relative to the $\etac (1S)$ meson are 
\begin{align*}
\frac{\BR ( b \to \chiczero X )}{\BR ( b \to \etac (1S) X )} &= 0.62 \pm 0.10 \pm 0.05 \pm 0.15_{\BR} , \\
\frac{\BR ( b \to \chicone X )}{\BR ( b \to \etac (1S) X )}  &= 0.56 \pm 0.12 \pm 0.05 \pm 0.13_{\BR} , \\
\frac{\BR ( b \to \chictwo X )}{\BR ( b \to \etac (1S) X )}  &= 0.23 \pm 0.04 \pm 0.02 \pm 0.06_{\BR} , 
\end{align*}
where the dominating uncertainty is due to the uncertainty of the 
branching fractions $\BR ( \etac (1S) \to \phi \phi )$ and $\BR ( \chi_c \to \phi \phi )$. 
The absolute branching fractions are determined to be 
\begin{align*}
\BR ( b \to \chiczero X ) &= ( 3.02 \pm 0.47 \pm 0.23 \pm 0.94_{\BR} ) \times 10^{-3} , \\
\BR ( b \to \chicone X )  &= ( 2.76 \pm 0.59 \pm 0.23 \pm 0.89_{\BR} ) \times 10^{-3} , \\
\BR ( b \to \chictwo X )  &= ( 1.15 \pm 0.20 \pm 0.07 \pm 0.36_{\BR} ) \times 10^{-3} , 
\end{align*}
where the third uncertainty is due to the uncertainties on the branching fractions of the \bquark-hadron decays to the $\etac (1S)$ meson, 
$\BR ( \bquark \to \etac (1S) X )$, and of $\etac (1S)$ and $\chi_c$ decays to $\phi \phi$. 
The branching fraction of \bquark-hadron decays into \chiczero is measured for the first time, 
and is found to be larger than the values predicted in Ref.~\cite{Beneke:1998ks}. 

Throughout the paper comparisons of the results on the production of charmonium states to theory predictions neglect the fact 
that the measured branching fractions also contain decays via intermediate higher-mass charmonium resonances, 
whereas theory calculations consider only direct \bquark-hadron transitions to the charmonium state considered.
Among the contributions that can be quantified the most sizeable comes from the \psitwos state decaying to the $\chi_c$ states.
With the branching fraction $\BR ( \decay{\bquark}{\psitwos X} )$ recently measured~\cite{LHCb-PAPER-2011-045} by \lhcb 
the branching fractions $\BR ( \decay{\bquark}{\chi_c X} )$, measured in this paper, are influenced by about $10 \%$. 
The branching fractions $\BR ( \decay{\bquark}{\chiczero X} )$  and $\BR ( \decay{\bquark}{\chictwo X} )$ 
remain different from the predictions in Ref.~\cite{Beneke:1998ks}.

The branching fraction measurement for \bquark-hadron decays into \chicone is most precise in mixtures of \Bz, \Bp, \Bs, \Bcp and \bquark baryons. 
The central value is lower than the central values measured by the DELPHI~\cite{Abreu:1994rk} and L3~\cite{Adriani:1993ta} 
experiments at LEP, $( 11.3 ^{+ 5.8} _{- 5.0} \pm 0.4 ) \times 10^{-3}$ and $( 19 \pm 7 \pm 1 ) \times 10^{-3}$, respectively. 
For the measurements with different \bquark-hadron content, the \lhcb result is consistent with 
measurements by \cleo~\cite{Anderson:2002md}, \belle~\cite{Bhardwaj:2015rju}, and \babar~\cite{Aubert:2002hc}. 
Finally, the \lhcb result for the inclusive \bquark-hadron decays into \chicone is consistent with the prediction in Ref.~\cite{Beneke:1998ks}. 

The branching fraction of \bquark-hadron decays into \chictwo is measured for the first time with a mixture of \Bz, \Bp, \Bs, \Bcp 
and \bquark-baryons.  
The result is consistent with the world average~\cite{PDG2016} measured with the \Bz and \Bp mixture, 
and with individual results from \cleo~\cite{Chen:2000ri}, \belle~\cite{Bhardwaj:2015rju} and \babar~\cite{Aubert:2002hc}. 
The value obtained is below the range predicted in Ref.~\cite{Beneke:1998ks}. 

\begin{figure}[t]
\centering
\begin{picture}(365,225)
\put(20,10){\includegraphics[width=360px]{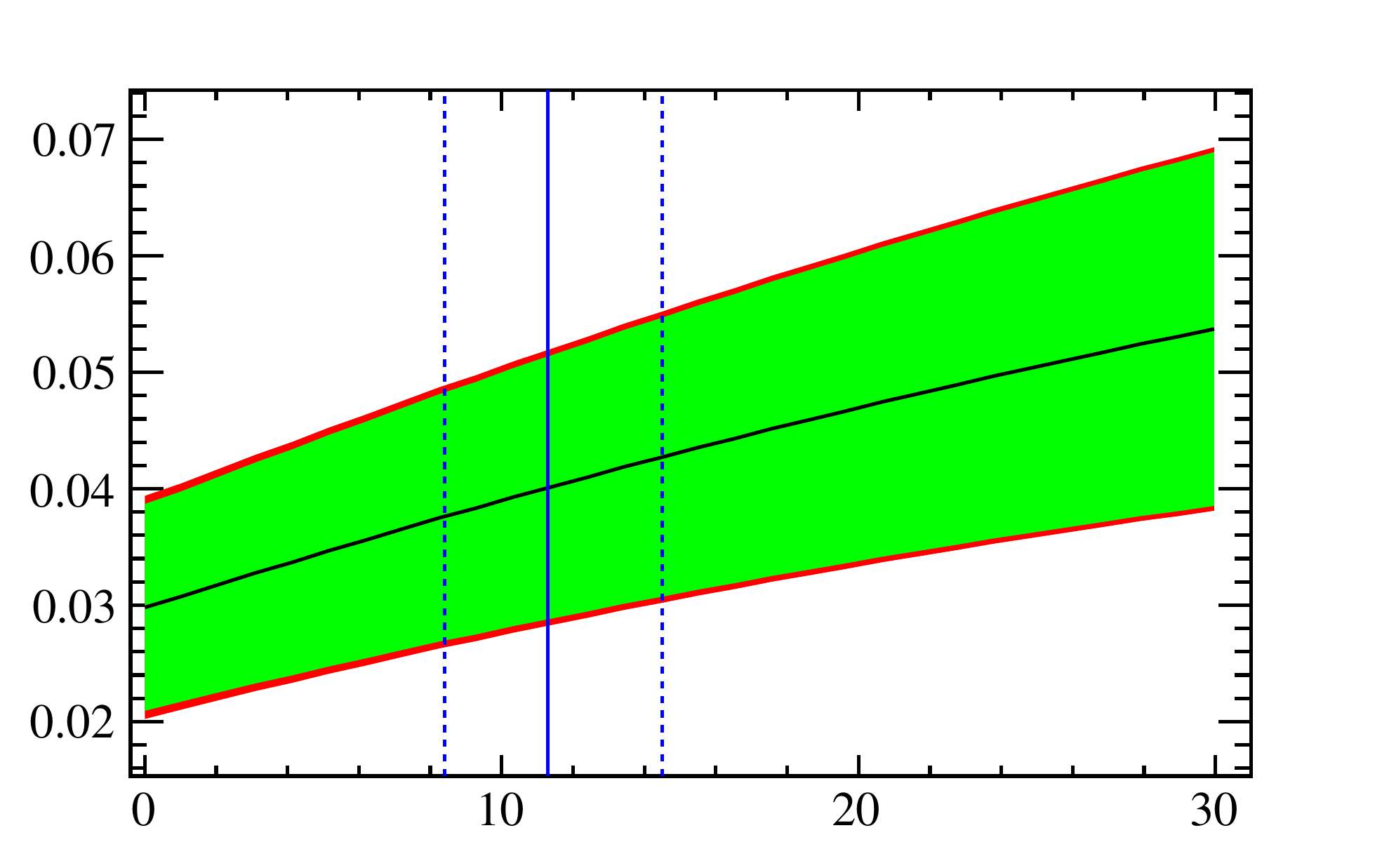}
                \put(-380,80){\rotatebox{90}{{$\frac{\BR ( b \to \etac (2S) X ) \times \BR ( \etac (2S) \to \phi \phi )}{\BR ( b \to \etac (1S) X ) \times \BR ( \etac (1S) \to \phi \phi )}$}}}
                \put(-310,180) {\lhcb}
                \put(-125,-2) {{$\Gamma ( \etac (2S) ) \ [ \mev ]$}}}
\end{picture}
\caption{Ratio of the $\etac(2S)$ and $\etac(1S)$ inclusive yields 
$\frac{\BR ( b \to \etac (2S) X ) \times \BR ( \etac (2S) \to \phi \phi )}{\BR ( b \to \etac (1S) X ) \times \BR ( \etac (1S) \to \phi \phi )}$
as a function of the assumed $\etac(2S)$ natural width. 
Statistical (green band) and total uncertainties are shown separately. 
The $\etac(2S)$ natural width from Ref.~\cite{PDG2016} is shown as a vertical solid line; the dashed lines indicate its uncertainty.
} \label{fig:etacvsgamma}
\end{figure}

A deviation of the $\etac(2S)$ natural width from the world average value~\cite{PDG2016} would affect the measured ratio 
of $\etac (2S)$ and $\etac (1S)$ production rates in \bquark-hadron inclusive decays, as shown in Fig.~\ref{fig:etacvsgamma}. 
The decay $\etac (2S) \to \phi \phi$ has not been observed so far. Hence the product of the branching fraction of \bquark-hadron decays 
to $\etac (2S)$ 
and the branching fraction of the $\etac (2S) \to \phi \phi$ decay mode is determined as
\begin{align*}
\BR ( b \to \etac (2S) X ) \times \BR ( \etac (2S) \to \phi \phi )
 &= ( 6.34 \pm 1.81 \pm 0.57 \pm 1.89 ) \times 10^{-7} , 
\end{align*}
where the systematic uncertainty is dominated by the uncertainty on the $\etac (1S)$ production rate in \bquark-hadron decays. 
This is the first evidence for $\etac (2S)$ production in \bquark-hadron decays, and for the decay of the $\etac (2S)$ meson into a pair of $\phi$ mesons.

\protect\subsection{Transverse momentum dependence of the differential cross-sections for $\boldsymbol{\etac (1S)}$ and $\boldsymbol{\chi_c}$ production}

\begin{figure}[t]
\centering
\begin{picture}(450,390)
\put(0,10){\includegraphics[width=450px]{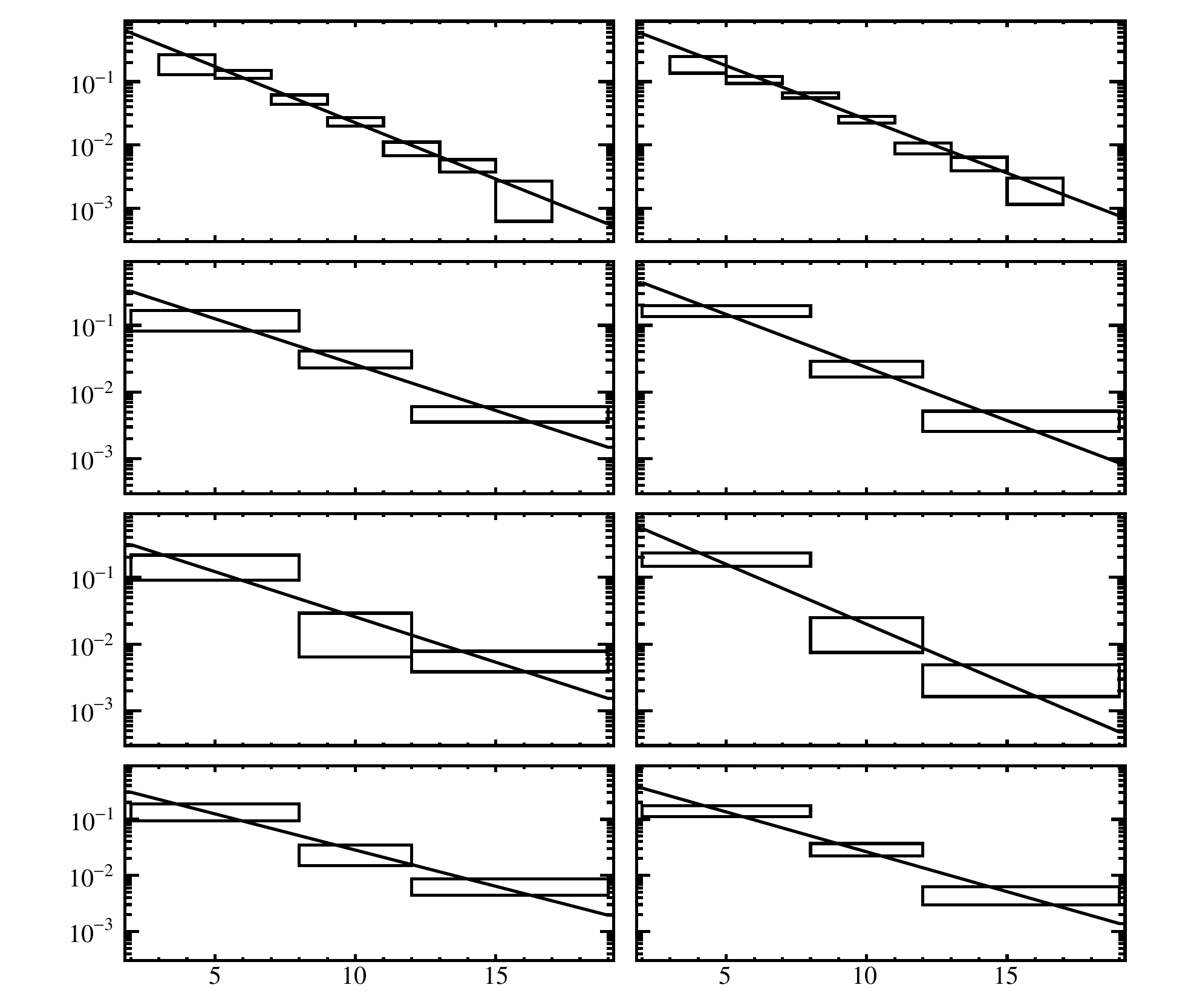}
                \put(-450,300){\rotatebox{90}{{$\frac{d \sigma}{\sigma^* dp_{\textrm{\tiny{$_{\mbox{T}}$}}}} \ [ \gev^{-1} ]$}}}
                \put(-395,300){a) $\etac (1S)$}
                \put(-395,205){c) \chiczero}
                \put(-395,110){e) \chicone}
                \put(-395,30){g) \chictwo}
                \put(-200,300){b) $\etac (1S)$}
                \put(-200,205){d) \chiczero}
                \put(-200,110){f) \chicone}
                \put(-200,30){h) \chictwo}
                \put(-270,360){\lhcb}
                \put(-270,265){\lhcb}
                \put(-270,170){\lhcb}
                \put(-270,75){\lhcb}
                \put(-270,345){$7 \tev$}
                \put(-270,250){$7 \tev$}
                \put(-270,155){$7 \tev$}
                \put(-270,60){$7 \tev$}
                \put(-70,360){\lhcb}
                \put(-70,265){\lhcb}
                \put(-70,170){\lhcb}
                \put(-70,75){\lhcb}
                \put(-70,345){$8 \tev$}
                \put(-70,250){$8 \tev$}
                \put(-70,155){$8 \tev$}
                \put(-70,60){$8 \tev$}
                \put(-75,-3){{$\pt \ [ \gev ]$}}}
\end{picture}
\protect\caption{Differential cross-sections normalized to the production cross-section integrated over the studied region, $\sigma^*$, 
of the (top to bottom) $\protect\etac (1S)$, \chiczero, \chicone and \chictwo states 
for the (left) $\protect\sqs = 7 \protect\tev$ and the (right) $\protect\sqs = 8 \protect\tev$ data samples. 
The horizontal and vertical size of the boxes reflect the size of the \pt bins and the statistical and uncorrelated systematic uncertainties 
of the differential production cross-sections added in quadrature. 
The exponential functions proportional to $\exp (- \alpha \, \pt )$ fitted to the integral of the each bin of the distributions are overlaid.
}
\label{fig:ptccbar}
\end{figure}
The shapes of the differential production cross-sections as a function of transverse momentum are studied 
in the \lhcb acceptance ($2 < \eta < 5$) and for $3 < \pt < 17 \gev$ and $2 < \pt < 19 \gev$ 
for the $\etac (1S)$ and $\chi_c$ states, respectively. 
Each differential production cross-section is normalized to the production cross-section integrated over the studied \pt region. 
Figure~\ref{fig:ptccbar} shows the normalized differential cross-sections of $\etac (1S)$, \chiczero, \chicone and \chictwo production at $\sqs = 7$ and $8 \tev$. 
An exponential function proportional to $\exp (- \alpha \, \pt )$ 
is fitted to the integral of the each bin of the distributions.
No significant difference is observed between the $\sqs = 7 \tev$ and $8 \tev$ data. 
The results for the slope parameters $\alpha$ are given in Table~\ref{tab:ptall}.
For \chicone and \chictwo production in \bquark-hadron decays these results extend the \atlas studies~\cite{ATLAS:2014ala} 
in \pt and rapidity.
\begin{table}[t]
\centering
\protect\caption{Exponential slope parameter in units of $\gev^{-1}$ from a fit to the \pt spectra of 
$\etac (1S)$, \chiczero, \chicone and \chictwo mesons. 
\protect\label{tab:ptall}}
\begin{tabular}{l|c|c|c|c}
              & $\etac (1S)$ & \chiczero & \chicone & \chictwo \\ \hline
$\sqs = 7 \tev$ & $0.41 \pm 0.02$ & $0.32 \pm 0.04$ & $0.31 \pm 0.06$ & $0.30 \pm 0.05$  \\ 
$\sqs = 8 \tev$ & $0.39 \pm 0.02$ & $0.37 \pm 0.04$ & $0.41 \pm 0.06$ & $0.33 \pm 0.04$ 
\end{tabular}
\end{table}

\subsection{Search for production of $\boldsymbol{X(3872)}$, $\boldsymbol{X(3915)}$ and $\boldsymbol{\chictwo (2P)}$}

The observation of the $X(3915)$ and $\chictwo (2P)$ states in \bquark-hadron decays or the $X(3872)$ decaying to a pair of $\phi$ mesons
would provide interesting information on the properties of these states. 
The invariant mass spectrum of $\phi \phi$ combinations in Fig.~\ref{fig:ccphiphi} shows no evidence for a signal 
from the $X(3872)$, $X(3915)$, or $\chictwo (2P)$ states. 
Bayesian upper limits assuming a uniform prior in the event yields are obtained 
on the branching fractions relative to those involving decays to the states with similar quantum numbers. 
For the states with similar quantum numbers, in the efficiency ratios systematic uncertainties largely cancel. 
Using the efficiency ratios from the simulation, the upper limits at $95 \%$ ($90 \%$) CL on the ratios of inclusive branching fractions are 
\begin{align*}
R^{X(3872)}_{\chicone} & < 0.39 \ (0.34) , \\
R^{X(3915)}_{\chiczero} & < 0.14 \ (0.12) , \\ 
R^{\chictwo (2P)}_{\chictwo} & < 0.20 \ (0.16) .
\end{align*}
Using the measured production rates of the $\chi_c$ states in \bquark-hadron decays and branching fractions for the $\chi_c$ decays to the $\phi \phi$ final state~\cite{PDG2016}, 
the upper limits at $95\%$ ($90 \%$) CL on the production rates of the $X(3872)$, $X(3915)$, and $\chictwo (2P)$ states in \bquark-hadron decays are 
\begin{align*}
\BR ( \bquark \ra X(3872) X ) \times \BR ( X(3872) \ra \phi \phi ) & < 4.5 \ (3.9) \times 10^{-7} , \\
\BR ( \bquark \ra X(3915) X ) \times \BR ( X(3915) \ra \phi \phi ) &  < 3.1 \ (2.7) \times 10^{-7} , \\
\BR ( \bquark \ra \chictwo (2P) X ) \times \BR ( \chictwo (2P) \ra \phi \phi ) &  < 2.8 \ (2.3)  \times 10^{-7} .
\end{align*}

\section{Masses and natural widths of charmonium states}
\label{sec:masses}

The majority of the $\etac (1S)$ mass measurements, used in the fit of Ref.~\cite{PDG2016}, were performed 
with two-photon production, $\gamma \gamma \to \etac (1S) \to \mbox{hadrons}$, 
radiative decays $\jpsi \to \etac (1S) \gamma$ and $\psitwos \to \etac (1S) \gamma$, 
$\proton \antiproton \to \etac (1S) \to \gamma \gamma$, and exclusive \B decays, yielding the average value $2983.4 \pm 0.5 \mev$. 
Mass determinations via exclusive $B$ decays, performed at the \babar and \belle experiments~\cite{Wu:2006vx,Aubert:2008kp,Vinokurova:2011dy},
do not provide consistent results. 
In 2009, the \cleo collaboration observed a significant asymmetry in the line shapes of radiative 
$\jpsi \to \gamma \etac (1S)$ and $\psitwos \to \gamma \etac (1S)$ transitions~\cite{Mitchell:2008aa},
which, when ignored, could lead to significant bias in the mass and width measurement via \jpsi or \psitwos radiative decays. 
Recent BES~III results~\cite{BESIII:2011ab,Ablikim:2012ur} 
obtained using radiative decays of \psitwos, shifted the world average value by more than two standard deviations. 
Therefore precise $\etac (1S)$ mass measurements using a different technique are needed. 
\lhcb measured \mbox{$M_{\etac (1S)} = 2982.2 \pm 1.5 \pm 0.1 \mev$}~\cite{LHCb-PAPER-2014-029} using 
$\etac (1S)$ from \bquark-hadron decays and reconstructing $\etac (1S)$ via decays to $\proton \antiproton$.
A similar situation occurs with the $\etac (1S)$ natural width determination, where
recent BES~III results
obtained using radiative decays of \psitwos 
shifted the world average from $29.7 \pm 1.0 \mev$ to $31.8 \pm 0.8 \mev$. 

The properties of the $\etac (2S)$ state are less well studied. 
Measurements at the \cleo~\cite{Asner:2003wv}, \babar~\cite{Aubert:2005tj,delAmoSanchez:2011bt}, \belle~\cite{Abe:2007jna,Vinokurova:2011dy} 
and BES~III~\cite{Ablikim:2013gd,Ablikim:2012sf} experiments, 
using $\gamma \gamma \to \etac (2S) \to \mbox{hadrons}$, double charmonium production in \epem annihilation, exclusive \B decays and radiative transitions of \psitwos,
yield the world averages~\cite{PDG2016} 
of $3639.4 \pm 1.3 \mev$ for the $\etac(2S)$ mass, and $11.3 ^{+ 3.2} _{-2.9} \mev$ for its natural width. 

Table~\ref{tab:mres} presents measurements of the masses of the $\etac$ and $\chi_c$ states and of the natural width of the $\etac (1S)$ 
from the fit of the $\phi \phi$ invariant mass spectrum in Fig.~\ref{fig:ccphiphi}. 
For the determination of the systematic uncertainties, except for the test of the impact of the $f_0 (980)$ meson, 
the same variations of the analysis are performed as for the determination of the charmonium yields. 
In addition, the effect of excluding the $\etac (2S)$ mass region ($2.8 - 3.7 \gev$) is studied, and the uncertainties
related to the momentum-scale calibration are estimated by varying the calibration parameter by $\pm \, 3 \times 10^{-4}$~\cite{LHCb-PAPER-2012-048}. 
The resulting total systematic uncertainty is obtained as the quadratic sum of the individual contributions. 
The uncertainty related to the momentum-scale calibration dominates the mass determination for all $\etac$ and $\chi_c$ states. 
The uncertainty of the $\Gamma_{\etac (1S)}$ measurement is dominated by the background description.
\addtolength{\tabcolsep}{-4pt}
\begin{table}[b]
\centering
\caption{Charmonium masses and natural widths in$\, \! \mev$. 
\label{tab:mres}}
\begin{tabular}{l|r|rcl}
                      & Measured value $\ \ $           & \multicolumn{3}{c}{$\ $ World average~\cite{PDG2016}} \\ \hline
$M_{\etac(1S)} \ $     & $\ 2982.8 \pm 1.0 \pm 0.5 \ $   & 2983.4 & $\pm$ & 0.5     \\ 
$M_{\chiczero}$        & $3413.0 \pm 1.9 \pm 0.6 \ $        & $\ $ 3414.75 & $\pm$ & 0.31  \\ 
$M_{\chicone}$         & $3508.4 \pm 1.9 \pm 0.7 \ $        & 3510.66 & $\pm$ & 0.07  \\ 
$M_{\chictwo}$         & $3557.3 \pm 1.7 \pm 0.7 \ $        & 3556.20 & $\pm$ & 0.09   \\ 
$M_{\etac(2S)}$        & $3636.4 \pm 4.1 \pm 0.7 \ $        & 3639.2 & $\pm$ & 1.2     \\ 
$\Gamma_{\etac(1S)}$   & $31.4 \pm 3.5 \pm 2.0 \ $          & 31.8 & $\pm$ & 0.8      \\ 
$\Gamma_{\etac(2S)}$   & $ - \ \ \ \ \ \ \ \ \ \ \ $          & \multicolumn{3}{c}{$11.3^{\, \ + \, \ 3.2}_{\, \ - \, \ 2.9} \ $}
\end{tabular}
\end{table}
\addtolength{\tabcolsep}{4pt}

The measured charmonium masses agree with the world averages~\cite{PDG2016}. 
The measured $\etac (1S)$ mass is in agreement with the previous \lhcb measurement 
using decays to the $\proton \antiproton$ final states~\cite{LHCb-PAPER-2014-029} and has a better precision. 
The precision obtained for the $\etac (1S)$ mass is comparable to the precision of the world average value. 
The value of the $\etac (1S)$ natural width is consistent with the world average~\cite{PDG2016}. 

\addtolength{\tabcolsep}{-4pt}
\begin{table}[t]
\centering
\caption{Charmonium mass differences (in$\, \! \mev$). 
\label{tab:dmres}}
\begin{tabular}{l|r|rcl}
                               & Measured value $\ $         & \multicolumn{3}{c}{$\ $ World average~\cite{PDG2016}} \\ \hline
$M_{\chicone} - M_{\chiczero}$   & $95.4 \pm 2.7 \pm 0.1 \ $      & 95.91 & $\pm$ & 0.83 \\ 
$M_{\chictwo} - M_{\chiczero}$     & $\ 144.3 \pm 2.6 \pm 0.2 \ $ & 141.45 & $\pm$ & 0.32 \\ 
$M_{\etac(2S)} - M_{\etac(1S)} \ $    & $653.5 \pm 4.2 \pm 0.4 \ $      & $\ $ 655.70 & $\pm$ & 1.48  
\end{tabular}
\end{table}
\addtolength{\tabcolsep}{4pt}
The charmonium mass differences $M_{\chicone} - M_{\chiczero}$, $M_{\chictwo} - M_{\chiczero}$, and $M_{\etac(2S)} - M_{\etac(1S)}$
are obtained (Table~\ref{tab:dmres}) as a consistency check and for comparison with theory.
For the determination of the systematic uncertainties the same variations of the analysis are performed as for the determination 
of the charmonium masses and widths. 
The uncertainty related to the 2D fit dominates the $M_{\chicone} - M_{\chiczero}$ mass difference measurement. 
The systematic uncertainty of the  $M_{\chictwo} - M_{\chiczero}$ measurement is dominated by the uncertainty related to 
potential contributions from other resonances and by the uncertainty on the background model. 
The uncertainty related to the momentum-scale calibration dominates the $M_{\etac(2S)} - M_{\etac(1S)}$ mass difference measurement. 
The measured charmonium mass differences agree with the world averages. 

\begin{figure}[t]
\centering
\begin{picture}(350,260)
\put(0,10){\includegraphics[width=360px]{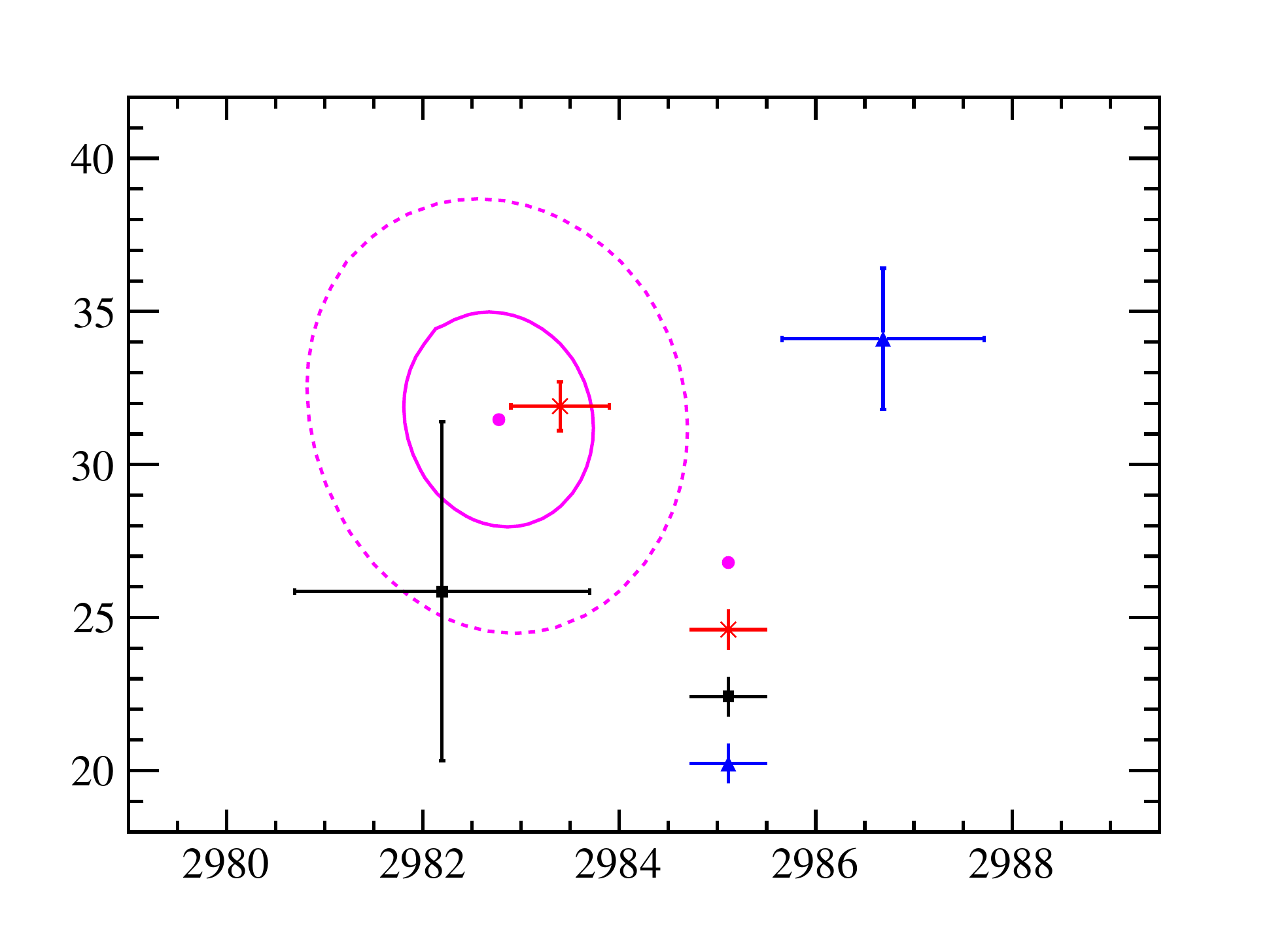}
                \put(-105,-3) {$M_{\etac (1S)} \ [ \mev ]$}
                \put(-140,100) {{\small{{This measurement}}}}
                \put(-140,81) {{\small{{World average~\cite{PDG2016}}}}}
                \put(-140,62) {{\small{{Inclusive \proton\antiproton~\cite{LHCb-PAPER-2014-029}}}}}
                \put(-140,43) {{\small{{Exclusive~\cite{LHCb-PAPER-2016-016}}}}}
                \put(-360,165){\rotatebox{90}{$\Gamma_{\etac (1S)} \ [ \mev ]$}}
                \put(-310,210) {\lhcb}
                \put(-200,202) {\small{$\Delta\chisq = 4$}}
                \put(-240,180) {\small{$\Delta\chisq = 1$}}}
\end{picture}
        \caption{Contour plot of $\Gamma_{\etac (1S)}$ and $M_{\etac (1S)}$ using $\etac \to \phi \phi$ decays. 
	The two magenta curves indicate $\Delta\chisq = 1$ and $\Delta\chisq = 4$ contours. 
	Only statistical uncertainties are shown. The red cross, black square and blue triangle with error bars indicate 
        the world average~\cite{PDG2016}, the result from Ref.~\cite{LHCb-PAPER-2014-029},
        and the result from Ref.~\cite{LHCb-PAPER-2016-016}, respectively.} 
\label{fig:contb}
\end{figure}
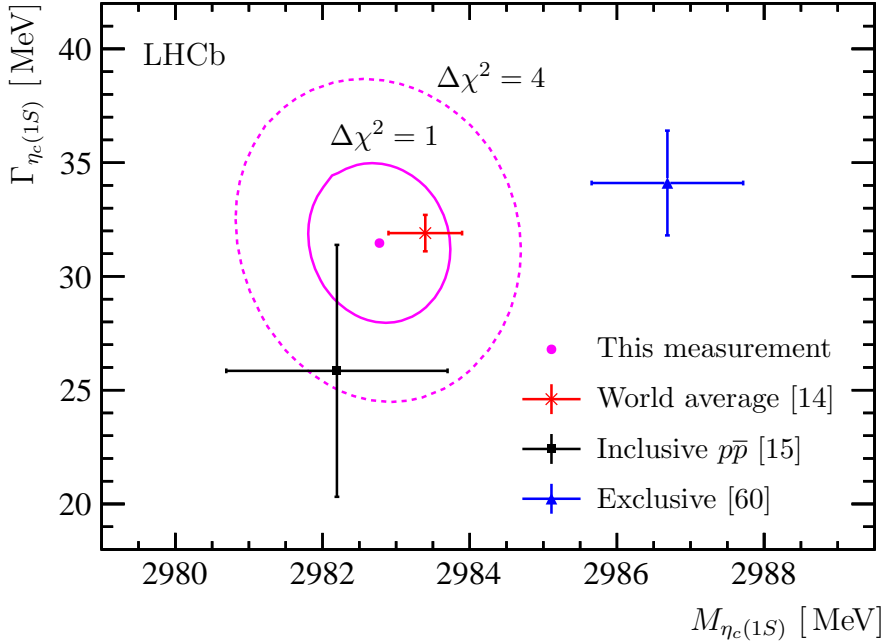

Figure~\ref{fig:contb} shows the $\Gamma_{\etac (1S)}, \, M_{\etac (1S)}$ contour plot, obtained from the analysis 
of \bquark-hadron decays into \etac mesons, where the \etac candidates are reconstructed via the decay $\etac (1S) \to \phi \phi$. 
The measurements of the $\etac (1S)$ mass and natural width using $\etac (1S)$ meson decays to $\phi \phi$ are consistent 
with the studies using decays to \proton\antiproton~\cite{LHCb-PAPER-2014-029} 
and with the world average~\cite{PDG2016}. 
The measured  $\etac (1S)$ mass is below the result in Ref.~\cite{LHCb-PAPER-2016-016}.
The precision obtained on the $\etac (1S)$ mass is comparable to the precision of the world average.

\section{First evidence of the $\boldsymbol{\Bs \to \phi \phi \phi}$ decay}
\label{sec:bs}

\begin{figure}[h]
\centering
\begin{picture}(370,225)
\put(20,10){\includegraphics[width=360px,height=8.cm]{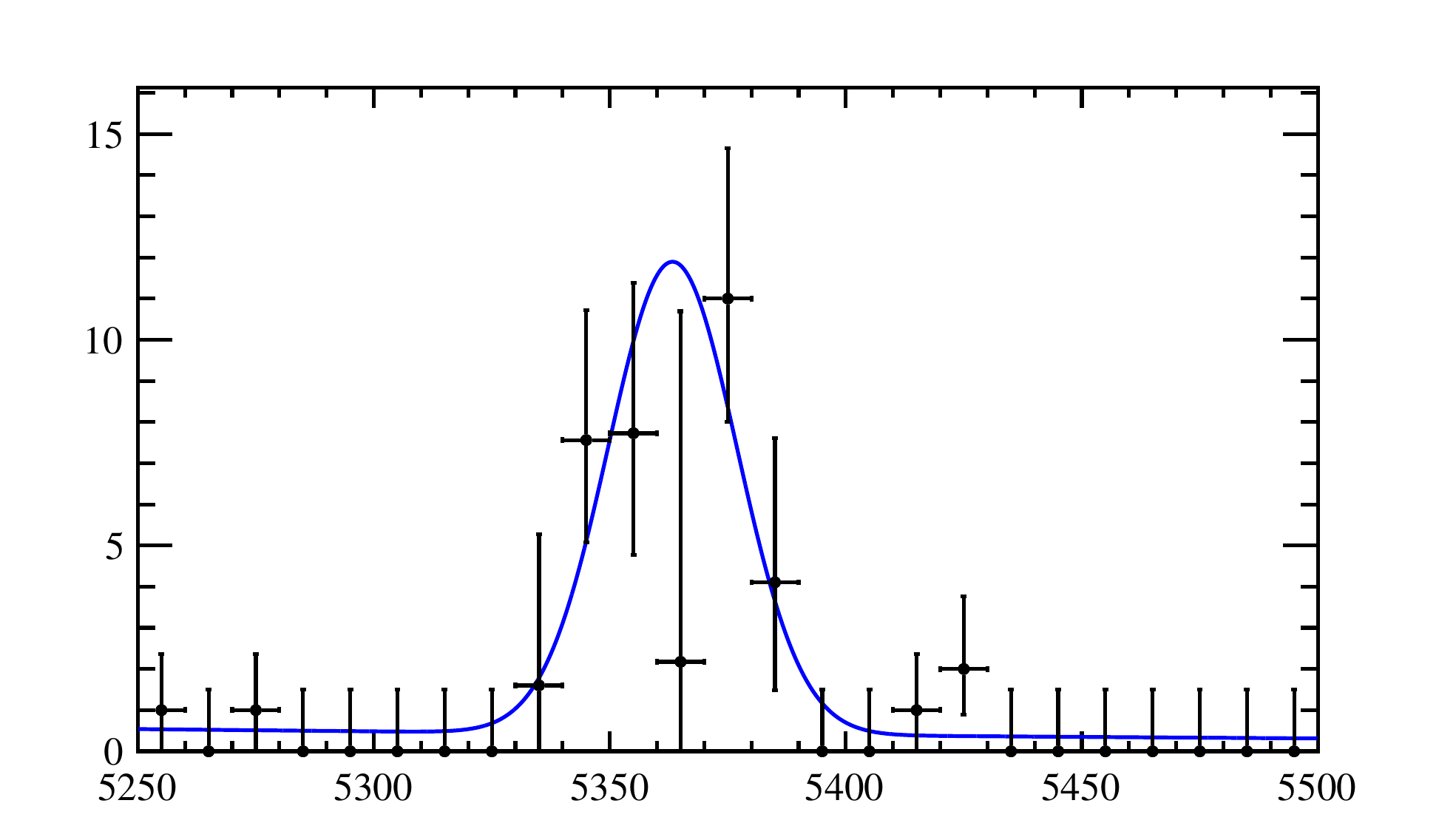}
                \put(-360,90){\rotatebox{90}{{Candidates/$( 10 \mev )$}}}
                \put(-120,-5) {{$M( \phi \phi \phi) \ [ \mev ]$}}
                \put(-90,180) {\lhcb}}
\end{picture}
\protect\caption{Invariant mass spectrum of the $\phi \phi \phi$ combinations in the region of the \Bs mass, 
including the fit function described in the text.} 
\label{fig:bstriphi}
\end{figure}
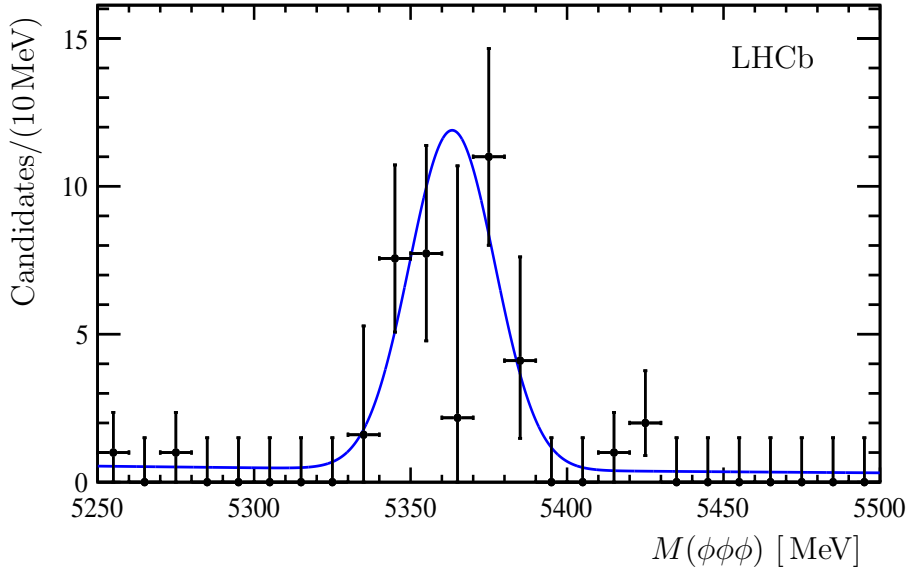

In order to extract $\phi \phi \phi$ combinations a 3D extended unbinned maximum likelihood fit is used, 
as described in Sect.~\ref{sec:sel}. 
Figure~\ref{fig:bstriphi} shows the invariant mass distribution for $\phi \phi \phi$ combinations. 
The fit to the invariant $\phi \phi \phi$ mass spectrum is performed using a sum of two Gaussian functions with a common mean to describe the \Bs signal, 
and an exponential function to describe combinatorial background. 
The ratio of the two Gaussian widths and the fraction of the narrow Gaussian are taken from simulation
so that a single free parameter in the $\phi \phi \phi$ invariant mass fit accounts for the detector resolution. 
A signal of $41 \pm 10 \pm 5$ \Bs decays over a low background of about 3 events is obtained. 
Uncertainties related to the background description in the 3D fit and to the decay model defining the $\phi$ polarization 
dominate the systematic uncertainty in the \Bs signal yield determination. 
The significance of the signal is estimated from the distributions of the difference in the logarithm of the best-fit \chisq
with and without including the signal shape 
in toy simulation samples. 
This leads to a signal significance of 4.9 standard deviations. 

\begin{figure}[t]
\centering
\begin{picture}(370,225)
\put(20,10){\includegraphics[width=360px,height=8.cm]{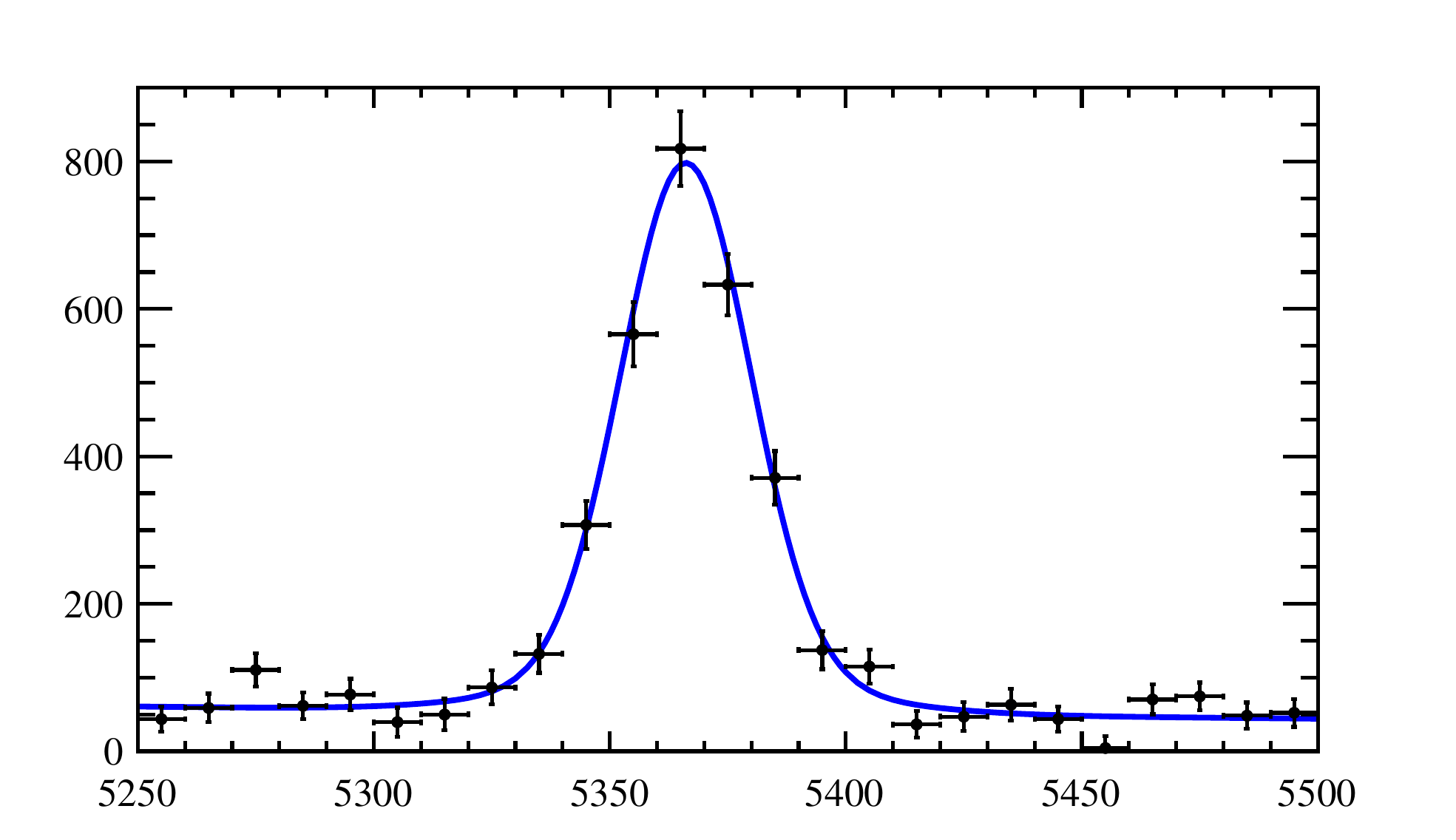}
                \put(-365,90){\rotatebox{90}{{Candidates/$( 10 \mev )$}}}
                \put(-115,-5) {{$M( \phi \phi ) \ [ \mev ]$}}
                \put(-90,180) {\lhcb}}
\end{picture}
\caption{Invariant mass spectrum of the $\phi \phi$ combinations in the region of the \Bs mass, 
including the fit function described in the text.} 
\label{fig:bsphiphi}
\end{figure}
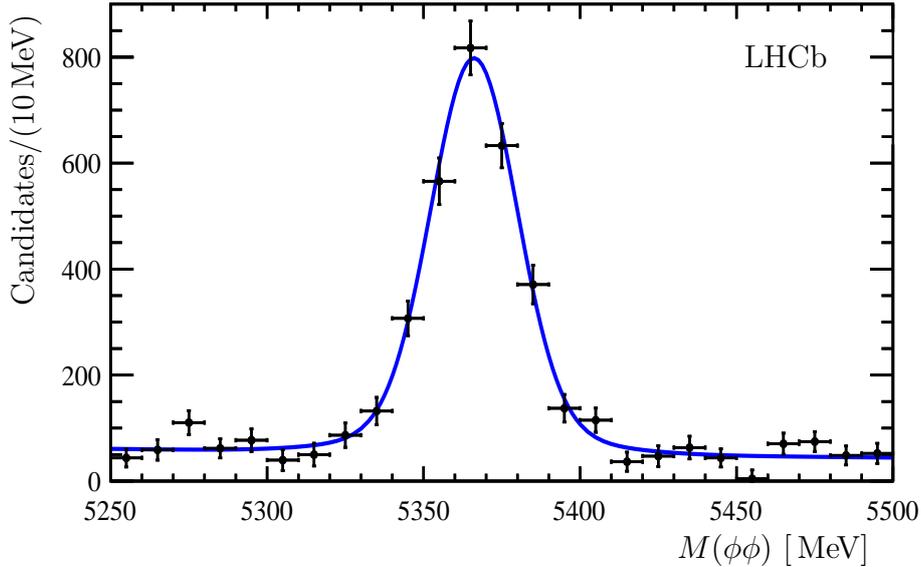

The $\Bs \to \phi \phi$ decay mode is chosen as a normalization mode for the $\BR ( \Bs \to \phi \phi \phi )$ measurement. 
The invariant mass spectrum obtained from 2D fits in bins of the $\phi \phi$ invariant mass in the region of the \Bs mass 
is shown in Fig.~\ref{fig:bsphiphi}. 
A sum of two Gaussian functions with a common mean is used to describe the \Bs signal shape, while an exponential function models the combinatorial background. 
The ratio of the two Gaussian widths and the fraction of the narrow Gaussian function are taken from simulation. 
In total $2701 \pm 114 \pm 84$ \Bs decays are found. 
The uncertainties related to the description of the resolution in the 2D fits and the description of the $\phi \phi$ invariant mass resolution
dominate the systematic uncertainty in the \Bs signal yield determination. 

The ratio of the $\Bs \to \phi \phi \phi$ and $\Bs \to \phi \phi$ branching fractions is obtained from 
the relative $\Bs \to \phi \phi \phi$ and $\Bs \to \phi \phi$ signal yields and their efficiencies as  
\begin{align*}
\frac{\BR ( \Bs \to \phi \phi \phi )}{\BR ( \Bs \to \phi \phi )} 
 = 
\frac{N_{\Bs \to \phi \phi \phi}}{N_{\Bs \to \phi \phi}}
 \times \frac{\varepsilon_{\Bs \to \phi \phi}}{\varepsilon_{\Bs \to \phi \phi \phi}} 
 \times \frac{1}{\BR ( \phi \to \Kp \Km )} 
 = 0.117 \pm 0.030 \pm 0.015 .
\end{align*}
In the above expression, the event yields are determined from the fits. 
The efficiency ratio,
$\varepsilon_{\Bs \to \phi \phi \phi} / \varepsilon_{\Bs \to \phi \phi} = 0.26 \pm 0.01$, 
is obtained from simulation and corrected to account for different \Bs transverse momentum spectra in data and simulation.
The $\Bs \to \phi \phi \phi$ transition is assumed to proceed via a three-body decay with uniform phase-space density. 
This assumption is supported below by comparing the $\phi \phi$ invariant mass distribution in data and simulation. 
The systematic uncertainty is dominated by the uncertainty in polarization of the $\phi$ mesons in the decay $\Bs \to \phi \phi \phi$, 
as discussed at the end of this section. 
Using the branching fraction of the $\Bs \to \phi \phi$ decay, 
$\BR ( \Bs \ra \phi \phi ) = (1.84 \pm 0.05 \pm 0.07 \pm 0.11_{f_s / f_d} \pm 0.12_{\scalebox{0.6}{norm}}) \times 10^{-5}$~\cite{LHCb-PAPER-2015-028}, 
the branching fraction for the \Bs meson decay to three $\phi$ mesons is determined to be 
\begin{align*}
\BR ( \Bs \to \phi \phi \phi ) = ( 2.15 \pm 0.54 \pm 0.28 \pm 0.21_{\BR} ) \times 10^{-6} ,
\end{align*}
where the last uncertainty is due to the branching fraction $\BR ( \Bs \to \phi \phi )$. 

\begin{figure}[t]
\protect\centering
\begin{picture}(330,225)
\put(10,10){\protect\includegraphics[width=315px]{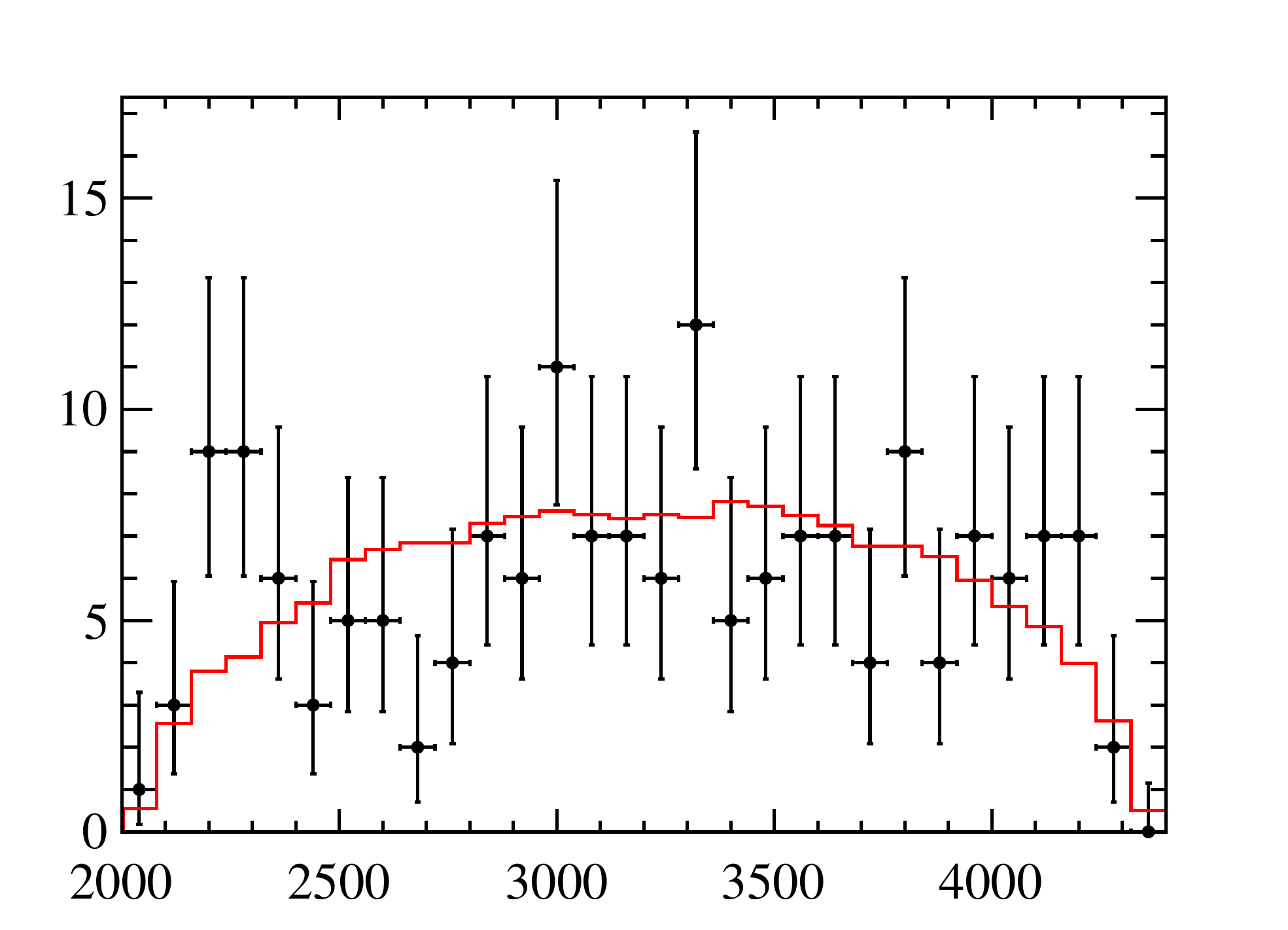}
                \protect\put(-320,95){\protect\rotatebox{90}{{Candidates/$( 80 \protect\mev )$}}}
                \protect\put(-105,-7) {{$M( \phi \phi) \ [ \mev ]$}}
                \protect\put(-270,185) {\protect\lhcb}}
\end{picture}
\protect\caption{The invariant mass distribution of each combination of $\phi \phi$ pairs in the $\protect\Bs \protect\to \phi \phi \phi$ candidates. 
The \Bs candidates are constrained to the known \Bs mass.
A phase-space distribution as obtained from simulation (red histogram) is overlaid.} 
\protect\label{fig:twotriphi}
\protect\end{figure}
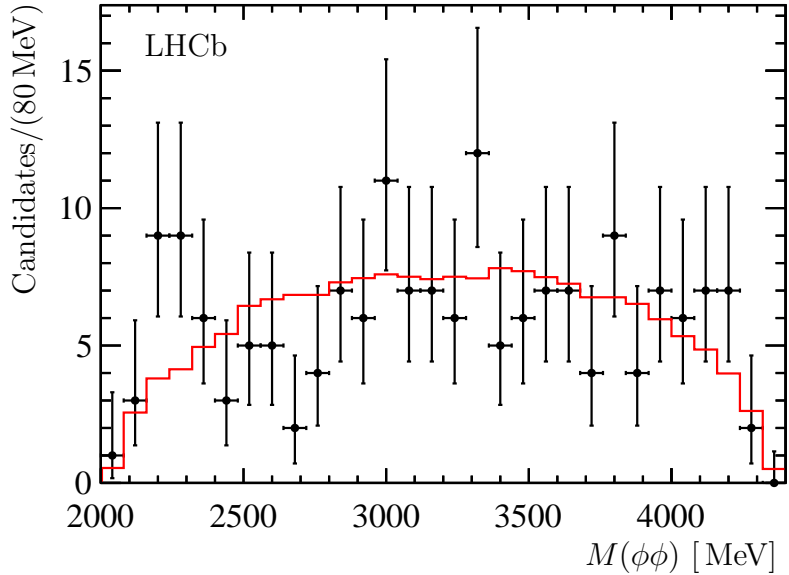

The $\Bs \to \phi \phi \phi$ transition can proceed via a two-body decay involving intermediate resonances 
or via a three-body $\Bs \to \phi \phi \phi$ decay.
In order to search for contributions from possible intermediate resonances, the invariant mass of each $\phi \phi$ combination 
from all $\Bs \to \phi \phi \phi$ candidates in the signal region of $\pm 3$ standard deviations around 
the \Bs mass is examined, see Fig.~\ref{fig:twotriphi}. 
The \Bs candidates are constrained to the known \Bs mass. 
Three entries to the histogram are produced by each \Bs candidate. 
A phase-space distribution as obtained from simulation is overlaid for comparison. 
No indication of significant contributions from \etac, $\chic$, $f_2 (2300)$ or $f_2 (2340)$ states is seen.
\protect
A symmetrized Dalitz plot constructed following the approach described in Ref.~\cite{Adolph:2008vn} 
shows no evidence for resonant contributions either. 

\begin{figure}[t]
\centering
\begin{picture}(330,225)
\put(10,10){\includegraphics[width=315px]{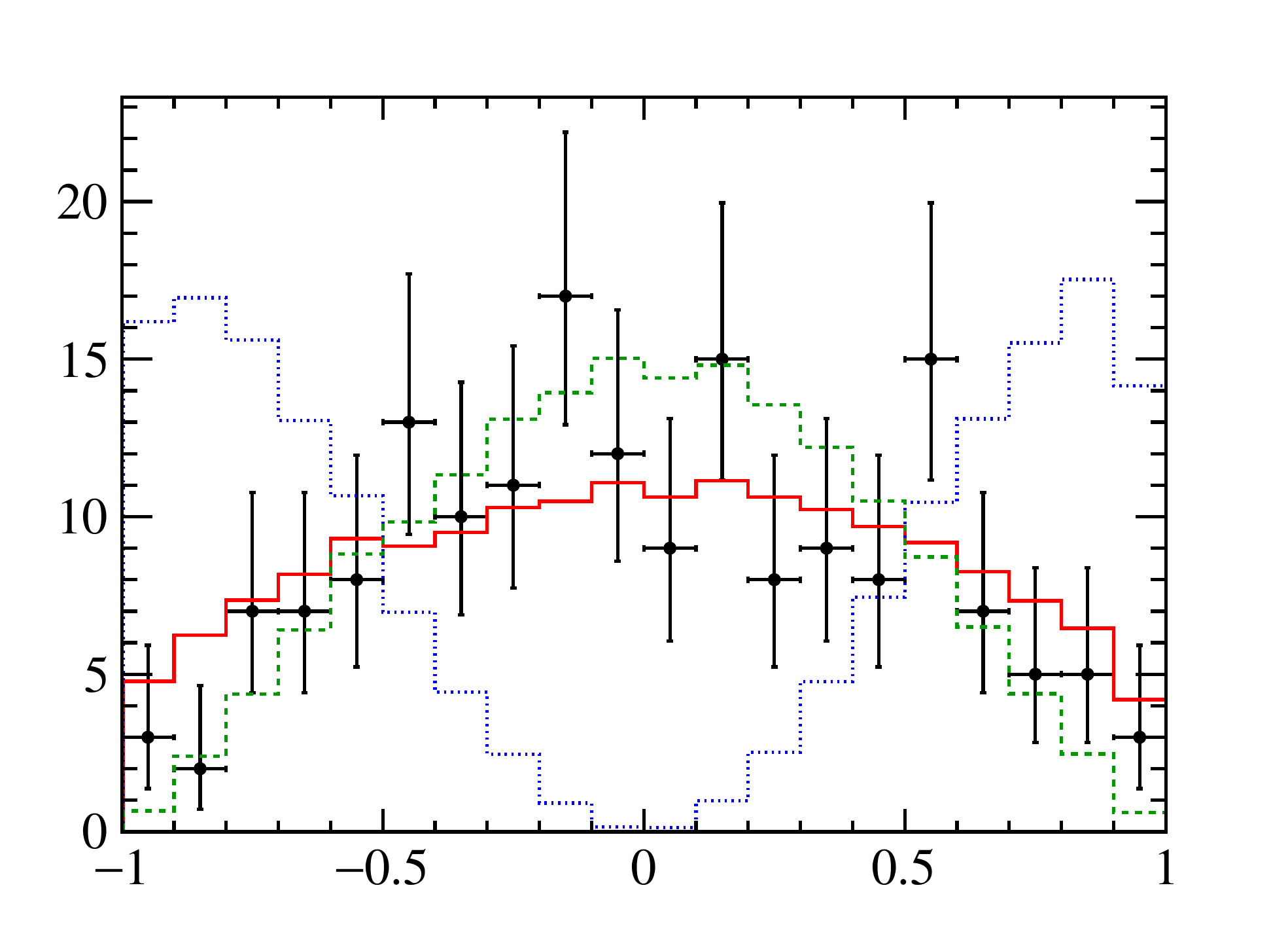}
                \put(-320,125){\rotatebox{90}{{Candidates/0.4}}}
                \put(-60,-3) {{$\cos ( \theta )$}}
                \put(-270,185) {\lhcb}}
\end{picture}
\protect\caption{The $\phi$ meson angular distribution for the $\protect\Bs \protect\to \phi \phi \phi$ candidates (points with error bars)
with the overlaid distribution from the simulation with no polarization (red solid histogram) and two extreme, transverse 
(green dashed histogram) and longitudinal (blue dotted histogram), polarizations.}
\protect\label{fig:pol}
\protect\end{figure}
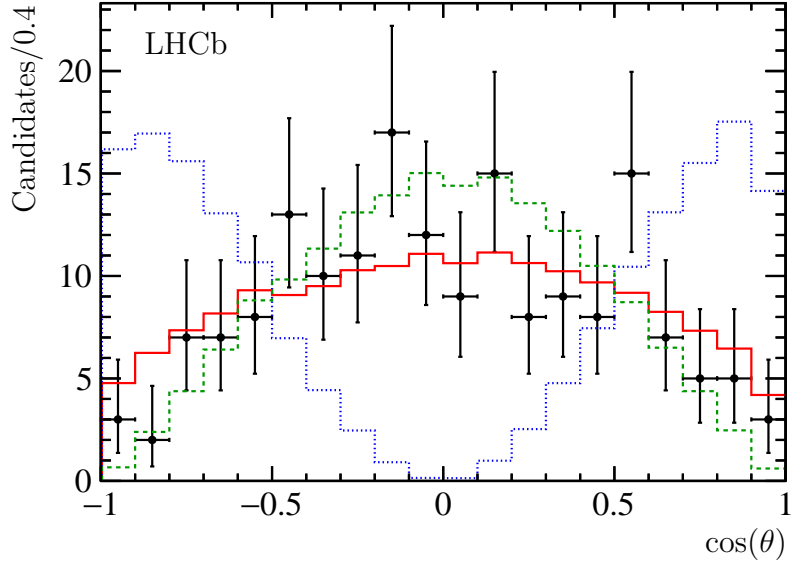

The polarization of the $\phi$ mesons is studied by means of the angle $\theta$ between the direction of flight of a $\phi$ meson 
in the \Bs rest frame and the \Bs direction in the laboratory frame. 
With the limited sample of $\Bs \to \phi \phi \phi$ candidates the 3D fit technique to remove contributions 
from $\Kp\Km$ combinations that are not from $\phi$ decays cannot be used for this measurement. 
Instead, all $\phi$ mesons contributing in the mass range of the \Bs are used, with an estimated signal purity of $71 \%$. 
Figure~\ref{fig:pol} 
compares the $\cos ( \theta )$ distribution for the $\Bs \to \phi \phi \phi$ signal candidates in data with expectations from simulation
using different assumptions for the polarization. 
The purely longitudinal polarization clearly does not describe the data. 
The difference between the expectations for no polarization and purely transverse polarization is used
to estimate the corresponding systematic uncertainty in the $\BR ( \Bs \to \phi \phi \phi )$ measurement. 
The most probable value for the fraction of transverse polarization, $f_{\textrm{T}}$, is found to be $f_{\textrm{T}} = 0.86$. 
Assuming a uniform prior in the physically allowed range, a Bayesian lower limit of $f_{\textrm{T}} > 0.28$ at $95\%$ CL is found. 
\protect

\protect\section{Summary and discussion}
\protect\label{sec:summary}

Charmonium production in \bquark-hadron inclusive decays is studied in \proton\proton collisions collected at $\sqs = 7$ and $8 \tev$ 
corresponding to an integrated luminosity of $3.0 \invfb$, using charmonium decays to $\phi$-meson pairs. 
The masses and natural widths of the \etac and $\chi_c$ states are determined. 
In addition, the first evidence of $\Bs \to \phi \phi \phi$ decay is obtained. 

Ratios of charmonium $C$ production rates, 
\begin{equation*}
    R^{C_1}_{C_2} \equiv \frac{\BR ( \bquark \to C_1\,X ) \times \BR ( C_1 \to \phi \, \phi )}{\BR ( \bquark \to C_2\,X ) \times \BR ( C_2 \to \phi \, \phi )} ,
\end{equation*}
are measured to be 
\begin{align*}
R^{\chiczero}_{\etac (1S)} &= 0.147 \pm 0.023 \pm 0.011 , \\
R^{\chicone}_{\etac (1S)} &= 0.073 \pm 0.016 \pm 0.006 , \\
R^{\chictwo}_{\etac (1S)} &= 0.081 \pm 0.013 \pm 0.005 , \\
R^{\chicone}_{\chiczero} &= 0.50 \pm 0.11 \pm 0.01 , \\
R^{\chictwo}_{\chiczero} &= 0.56 \pm 0.10 \pm 0.01 , \\
R^{\etac (2S)}_{\etac (1S)} &= 0.040 \pm 0.011 \pm 0.004 , 
\end{align*}
where the first uncertainties are statistical and the second ones are systematic. 
Using the branching fractions of $\chi_c$ decays to $\phi \phi$ from Ref.~\cite{PDG2016}, 
relative branching fractions of \bquark hadrons decaying inclusively to $\chi_c$ states are derived, 
\begin{align*}
\frac{\BR ( b \to \chicone X )}{\BR ( b \to \chiczero X )}  &= 0.92 \pm 0.20 \pm 0.02 \pm 0.14_{\BR} , \\
\frac{\BR ( b \to \chictwo X )}{\BR ( b \to \chiczero X )}  &= 0.38 \pm 0.07 \pm 0.01 \pm 0.05_{\BR} , 
\end{align*}
where the third uncertainty is due to the branching fractions $\BR ( \chi_c \to \phi \phi )$. 
These results are consistent with the ratio of the \chicone and \chictwo production rates measured in \Bz and \Bp decays~\cite{PDG2016}. 

Inclusive production rates of the $\chi_c$ states in \bquark-hadron decays are derived using 
branching fractions of the $\chi_c$ decays to $\phi \phi$ from Ref.~\cite{PDG2016}, 
an average of the results from \belle~\cite{Huang:2003dr} and \babar~\cite{Aubert:2004gc} 
$\BR ( \etac (1S) \to \phi \phi ) = ( 3.21 \pm 0.72 ) \times 10^{-3}$, 
and the $\etac (1S)$ inclusive production rate measured using decays to \proton\antiproton,
$\BR ( b \to \etac (1S) X ) = ( 4.88 \pm 0.97 ) \times 10^{-3}$~\cite{LHCb-PAPER-2014-029}. 
They are
\begin{align*}
\BR ( b \to \chiczero X ) &= ( 3.02 \pm 0.47 \pm 0.23 \pm 0.94_{\BR} ) \times 10^{-3} , \\
\BR ( b \to \chicone X )  &= ( 2.76 \pm 0.59 \pm 0.23 \pm 0.89_{\BR} ) \times 10^{-3} , \\
\BR ( b \to \chictwo X )  &= ( 1.15 \pm 0.20 \pm 0.07 \pm 0.36_{\BR} ) \times 10^{-3} , 
\end{align*}
where the third uncertainty is due to the uncertainties on the branching fraction of the \bquark-hadron decays to the $\etac (1S)$ meson, $\BR ( \bquark \to \etac (1S) X )$, 
and $\etac (1S)$ and $\chi_c$ decays to $\phi \phi$. 
No indirect contribution to the production rate is subtracted.
However, since contributions from \psitwos decays to the $\chi_c$ states are limited, the results disfavour dominance of either colour-octet 
or colour-singlet contributions. 
The observed relations between the $\chi_c$ branching fractions are not consistent with those predicted in Ref.~\cite{Beneke:1998ks}. 
The branching fraction $\BR ( b \to \chiczero X )$ is measured for the first time. 
The result for \bquark-hadron decays into \chicone is the most precise measurement for the mixture of \Bz, \Bp, \Bs, \Bcp and \bquark-baryons. 
The central value of the result for \bquark-hadron decays into \chicone is lower 
than the central values 
measured by the DELPHI~\cite{Abreu:1994rk} and L3~\cite{Adriani:1993ta} experiments at LEP. 
The value obtained is consistent with the branching fraction of \bquark-hadron decays into \chicone 
measured by \cleo~\cite{Anderson:2002md}, \belle~\cite{Bhardwaj:2015rju} and \babar~\cite{Aubert:2002hc} with the light mixture of \Bz and \Bp. 
The branching fraction of \bquark-hadron decays into \chictwo is measured for the first time with the \Bz, \Bp, \Bs and \bquark-baryons mixture.  
The result is consistent with the world average corresponding to the \Bz, \Bp mixture~\cite{PDG2016} 
and with individual measurements from \cleo~\cite{Chen:2000ri}, \belle~\cite{Bhardwaj:2015rju}, and \babar~\cite{Aubert:2002hc}. 

Scaled differential charmonium production cross-sections as a function of \pt are presented for the $\etac (1S)$ and $\chi_c$ states in the \lhcb acceptance and for $\pt > 4 \gev$. 
Next-to-leading-order calculations of the \pt dependence of the \etac and $\chi_c$ production rates in \bquark-hadron decays 
will help to relate the results to conclusions on production mechanisms. 

The production rate of the $\etac (2S)$ state in \bquark-hadron decays is determined to be  
\begin{align*}
\BR ( b \to \etac (2S) X ) \times \BR ( \etac (2S) \to \phi \phi )
 &= ( 6.34 \pm 1.81 \pm 0.57 \pm 1.89_{\BR} ) \times 10^{-7} .
\end{align*}
This is the first measurement for inclusive $\etac (2S)$ production rate in \bquark-hadron decays 
and the first evidence for the decay $\etac (2S) \to \phi \phi$. 
The production rate as a function of the assumed natural width is given in Fig.~\ref{fig:etacvsgamma}. 
These are the first $\chi_c$ and $\etac (2S)$ inclusive production measurements, using charmonium decays 
to a hadronic final state, in the high-multiplicity environment of a hadron machine. 
In addition, upper limits at $95 \%$ ($90 \%$) CL on the production rates of the $X(3872)$, $X(3915)$, and $\chictwo (2P)$ states 
in \bquark-hadron decays are obtained, 
\[ R^{X(3872)}_{\chicone} < 0.39 \ (0.34) , \ R^{X(3915)}_{\chiczero} < 0.14 \ (0.12) \ \mathrm{and} \  R^{\chictwo (2P)}_{\chictwo} < 0.20 \ (0.16) ,
\]
or 
\begin{align*}
\BR ( \bquark \ra X(3872) X ) \times \BR ( X(3872) \ra \phi \phi ) & < 4.5 (3.9) \times 10^{-7} , \\
\BR ( \bquark \ra X(3915) X ) \times \BR ( X(3915) \ra \phi \phi ) &  < 3.1 (2.7) \times 10^{-7} , \\
\BR ( \bquark \ra \chictwo (2P) X ) \times \BR ( \chictwo (2P) \ra \phi \phi ) &  < 2.8 (2.3)  \times 10^{-7} . 
\end{align*}

Masses and natural widths of the \etac and $\chi_c$ states agree with the world averages. 
The precision of the $\etac (1S)$ mass is comparable to the precision of the world average value. 
The measured $\etac (1S)$ mass is in agreement with the \lhcb measurement 
using decays to the $\proton \antiproton$ final states~\cite{LHCb-PAPER-2014-029}. 

First evidence for the transition $\Bs \to \phi \phi \phi$ is reported with a significance of 4.9 standard deviations. 
Its branching fraction is measured to be 
\begin{align*}
\BR ( \Bs \to \phi \phi \phi ) = ( 2.15 \pm 0.54 \pm 0.28 \pm 0.21_{\BR} ) \times 10^{-6} .
\end{align*} 
No resonant structure is observed in the $\phi \phi$ invariant mass distribution. 
In the $\Bs \to \phi \phi \phi$ decay, transverse polarization is preferred for the $\phi$ mesons, with an estimate of $f_{\textrm{T}} > 0.28$ at $95 \%$ CL 
 and the most probable value of $f_{\textrm{T}} = 0.86$ for the fraction of transverse polarization. 

As a by-product of the analysis, the branching fraction $\BR ( \Bs \to \phi \phi )$ is determined to be 
$\BR ( \Bs \to \phi \phi ) = ( 2.18 \pm 0.17 \pm 0.11 \pm 0.14_{f_s} \pm 0.65_{\BR} ) \times 10^{-5}$
with a different technique with respect to the previous results~\cite{Abe:1999ze,Acosta:2005eu,Aaltonen:2011rs,LHCb-PAPER-2015-028}. 
This technique is based on relation of \Bs production in \proton\proton collisions and $\etac (1S)$ inclusive production rate in \bquark-hadron decays,
and reconstruction of \Bs and $\etac (1S)$ via decays to $\phi \phi$. 
The measurement is consistent with the recent \lhcb result~\cite{LHCb-PAPER-2015-028} 
and the current world average~\cite{PDG2016}, 
as well as with theoretical calculations~\cite{Beneke:2006hg,Ali:2007ff,Cheng:2009mu}. 

Finally, using the measurements presented and external input, the ratio of the branching fractions 
for the $\etac (1S)$ decays to $\phi \phi$ and to $\proton \antiproton$ is determined. 
The measured \Bs and $\etac (1S)$ yields and efficiency ratio, the branching fraction 
\mbox{$\BR ( \Bs \ra \phi \phi ) = (1.84 \pm 0.05 \pm 0.07 \pm 0.11_{f_s / f_d} \pm 0.12_{\scalebox{0.6}{norm}}) \times 10^{-5}$}~\cite{LHCb-PAPER-2015-028}, 
the \jpsi production rate in \bquark-hadron decays $\BR ( \bquark \ra \jpsi X ) = (1.16 \pm 0.10 ) \%$~\cite{PDG2016}, 
the relative production rates of $\etac (1S)$ and \jpsi in \bquark-hadron decays 
$\frac{\BR( \bquark \ra \etac (1S) X) \times \BR( \etac (1S) \ra \proton \antiproton )}{\BR( \bquark \ra \jpsi X) \times \BR( \jpsi \ra \proton \antiproton )} = 0.302 \pm 0.042$~\cite{LHCb-PAPER-2014-029}, 
the branching fraction $ \BR( \jpsi \ra \proton \antiproton ) = ( 2.120 \pm 0.029 ) \times 10^{-3}$~\cite{PDG2016}, 
the ratio of fragmentation fractions \mbox{$f_s / f_d = 0.259 \pm 0.015$}~\cite{fsfd}, 
and the \Lb fragmentation fraction $f_{\Lb}$ momentum dependence from Ref.~\cite{LHCb-PAPER-2014-004} are used.
The ratio of the branching fractions for the $\etac (1S)$ decays to $\phi \phi$ and to $\proton \antiproton$ is determined as 
\[
\frac{\BR ( \etac (1S) \ra \phi \phi )}{\BR ( \etac (1S) \ra \proton \antiproton )} = 1.79 \pm 0.14 \pm 0.09 \pm 0.10_{f_s / f_d} \pm 0.03_{f_{\Lb}} \pm 0.29_{\BR} , 
\]
where the third uncertainty is related to $f_s / f_d$, the fourth uncertainty is related to $f_{\Lb}$, 
and the fifth uncertainty is related to uncertainties of the production rates and decay branching fractions involved. 
This value is larger than the value computed from the world average branching fractions given in Ref.~\cite{PDG2016}.


\section*{Acknowledgements}

\noindent We would like to thank Emi Kou for motivating the studies of charmonium production in \lhcb using hadronic final states 
and the useful discussions regarding charmonium production mechanisms. 
We express our gratitude to our colleagues in the CERN
accelerator departments for the excellent performance of the LHC. We
thank the technical and administrative staff at the LHCb
institutes. We acknowledge support from CERN and from the national
agencies: CAPES, CNPq, FAPERJ and FINEP (Brazil); MOST and NSFC (China);
CNRS/IN2P3 (France); BMBF, DFG and MPG (Germany); INFN (Italy);
NWO (The Netherlands); MNiSW and NCN (Poland); MEN/IFA (Romania);
MinES and FASO (Russia); MinECo (Spain); SNSF and SER (Switzerland);
NASU (Ukraine); STFC (United Kingdom); NSF (USA).
We acknowledge the computing resources that are provided by CERN, IN2P3 (France), KIT and DESY (Germany), INFN (Italy), SURF (The Netherlands), PIC (Spain), GridPP (United Kingdom), RRCKI and Yandex LLC (Russia), CSCS (Switzerland), IFIN-HH (Romania), CBPF (Brazil), PL-GRID (Poland) and OSC (USA). We are indebted to the communities behind the multiple open
source software packages on which we depend.
Individual groups or members have received support from AvH Foundation (Germany),
EPLANET, Marie Sk\l{}odowska-Curie Actions and ERC (European Union),
Conseil G\'{e}n\'{e}ral de Haute-Savoie, Labex ENIGMASS and OCEVU,
R\'{e}gion Auvergne (France), RFBR and Yandex LLC (Russia), GVA, XuntaGal and GENCAT (Spain), Herchel Smith Fund, The Royal Society, Royal Commission for the Exhibition of 1851 and the Leverhulme Trust (United Kingdom).



\addcontentsline{toc}{section}{References}
\setboolean{inbibliography}{true}
\bibliographystyle{LHCb}
\bibliography{main,LHCb-PAPER,LHCb-CONF,LHCb-DP,LHCb-TDR}

\newpage

\centerline{\large\bf LHCb collaboration}
\begin{flushleft}
\small
R.~Aaij$^{40}$,
B.~Adeva$^{39}$,
M.~Adinolfi$^{48}$,
Z.~Ajaltouni$^{5}$,
S.~Akar$^{59}$,
J.~Albrecht$^{10}$,
F.~Alessio$^{40}$,
M.~Alexander$^{53}$,
S.~Ali$^{43}$,
G.~Alkhazov$^{31}$
% ...

{\footnotesize \it
% ...
$ ^{5}$Clermont Universit{\'e}, Universit{\'e} Blaise Pascal, CNRS/IN2P3, LPC, Clermont-Ferrand, France\\
% ...
$ ^{10}$Fakult{\"a}t Physik, Technische Universit{\"a}t Dortmund, Dortmund, Germany\\
% ...
$ ^{31}$Petersburg Nuclear Physics Institute (PNPI), Gatchina, Russia\\
% ...
$ ^{39}$Universidad de Santiago de Compostela, Santiago de Compostela, Spain\\
$ ^{40}$European Organization for Nuclear Research (CERN), Geneva, Switzerland\\
% ...
$ ^{43}$Nikhef National Institute for Subatomic Physics, Amsterdam, The Netherlands\\
% ...
$ ^{48}$H.H. Wills Physics Laboratory, University of Bristol, Bristol, United Kingdom\\
% ...
$ ^{53}$School of Physics and Astronomy, University of Glasgow, Glasgow, United Kingdom\\
% ...
$ ^{59}$University of Cincinnati, Cincinnati, OH, United States\\
% ...
}
\end{flushleft}


%
%

\end{document}